\documentclass[11pt]{article}

\usepackage[american]{babel}
\usepackage{indentfirst}
\usepackage{amsmath}
\usepackage{graphicx}
\usepackage{tikz}
\usetikzlibrary{calc}
\usepackage{xcolor}
\usepackage{url}
\usepackage{mathptmx}
\usepackage[numbers,sort&compress]{natbib}
\usepackage[margin=1in]{geometry}
\usepackage{authblk}

\title{Numerical Integration Instabilities in Likelihood Fits: Mechanisms and Remedies}
\author[1]{Manqi~Zhu}
\author[2]{Gerry~Bauer}
\author[2]{Kai~Yi}
\author[2]{Xining~Wang\thanks{e-mail: xining.wang@cern.ch}}
\affil[1]{Honors College, Nanjing Normal University, Nanjing 210023, China}
\affil[2]{Ministry of Education Key Laboratory of NSLSCS, Institute of Physics Frontiers and Interdisciplinary Sciences, School of Physics and Technology, Nanjing Normal University, Nanjing 210023, China}

\date{}

\begin{document}
\maketitle

\begin{abstract}
Numerical instabilities in the normalization of probability density functions can produce artificially lower negative log-likelihood (NLL) values than the true minimum of a likelihood fit, even when the resulting fit provides a poorer description of the data. Motivated by anomalously low NLL values found in non-converged fits associated with a recent CMS search for fully charmed tetraquark candidates, we use a pseudo-dataset generated from a likelihood model based on that analysis to investigate how numerical integration can yield unreliable normalization results, how these errors affect likelihood minimization, and how they can be addressed. We find that, even at fixed integration tolerance, the stopping decision of a numerical integrator can change with the fit parameters, introducing discontinuities in the normalization integrals that propagate to the likelihood surface. Such discontinuities can trap the minimizer at a numerical step and yield an artificially lower NLL. Tightening the integration tolerance mitigates these numerical discontinuities, while adaptive integration methods, such as Gauss--Kronrod quadrature, can also improve the numerical stability. A more severe failure occurs when the integration grid cannot adequately resolve a sharply localized contribution. In this case, the integrator may satisfy its estimated-error criterion despite substantially underestimating the normalization, allowing the minimizer to enter a region with an artificially low NLL. Tightening integration tolerances or using alternative adaptive integration methods does not necessarily resolve sharply localized contributions. We demonstrate that a suitable variable transformation can improve their numerical resolution, yielding more accurate normalization integrals and restoring fit convergence.
\end{abstract}

\section{Introduction}
\label{sec:intro}

Maximum-likelihood fitting is widely used for parameter estimation in many fields. In many applications, the normalization integrals of probability density functions (PDFs) depend on the fit parameters and must be evaluated numerically to calculate the likelihood. Numerical inaccuracies in these integrals can distort the evaluated likelihood, potentially affecting parameter estimation and fit convergence.

Likelihood fitting is extensively used in experimental particle physics, and we draw on the case of fitting structures in the $J/\psi J/\psi$ mass spectrum reported by three LHC experiments~\cite{LHCb:2020bwg,ATLAS:2023bft,CMS:2023owd,CMS:2025fpt,CMS:2026fam}.
In particular, CMS has reported a family of fully charmed tetraquark candidates whose masses exhibit a Regge pattern, with possible connections to structures observed in the $J/\psi\phi$ spectrum~\cite{CMS:2023owd,CMS:2025fpt,CMS:2026fam,Zhu:2024swp,Zhou:2025nyj}.
Extracting resonance properties from these spectra involves complex likelihood fits with multiple contributions, as exemplified by the CMS $J/\psi J/\psi$ analysis~\cite{CMS:2023owd,CMS:2025fpt,CMS:2026fam}, which includes three interfering resonances and yields multiple fit solutions~\cite{Bai:2019jrb}.
The fits were performed using RooFit~\cite{Verkerke:2003ir}, with RooIntegrator1D for numerical normalization and MIGRAD in Minuit~\cite{James:1975dr} for minimization.
In maximum-likelihood fits, a lower negative log-likelihood (NLL) ordinarily indicates a better description of the data within the assumed model. However, during the development of this analysis, some fits returned NLL values substantially lower than those obtained from normally converged fits, despite failing standard convergence criteria.
Likelihood scans of these supposed minima exhibited discontinuities.
This behavior was traced back to problems with numerical integrations performed in the minimization algorithm and motivated a broader investigation into the conditions under which numerical integration can produce inaccurate normalization results, how the resulting errors affect likelihood minimization, and how such failures can be identified and corrected.

In this work, we investigate this problem using a pseudo-dataset based on the CMS analysis~\cite{CMS:2023owd,CMS:2025fpt}.
Other technical challenges, such as the estimation of large global significances, have been discussed elsewhere~\cite{Chen:2025sym} and are beyond the scope of this study.
We examine how the criteria used to terminate numerical integration affect the accuracy and continuity of the normalization integrals and the resulting likelihood, and show that tightening the integration tolerance or using adaptive integration methods can mitigate the numerical discontinuities.
We further investigate a more severe failure in which a sharply localized contribution is not adequately resolved by the integration grid, causing the normalization to be substantially underestimated despite satisfying the estimated-error criterion. By examining the minimization process, we trace how the resulting likelihood distortion leads the minimizer into an anomalous parameter region. Finally, we compare different numerical integration approaches and demonstrate that a suitable variable transformation can improve the normalization accuracy and restore fit convergence.

\section{Model and Fit Setup}
\label{sec:setup}

The numerical study considers the $J/\psi J/\psi$ mass spectrum, $m=m_{J/\psi J/\psi}$, in the range $6<m<15~\mathrm{GeV}$.
Following the CMS analysis~\cite{CMS:2022pas}, the model consists of four components: nonresonant single-parton scattering (NRSPS), double-parton scattering (DPS), a noninterfering threshold component $\mathrm{BW}_0$, and a coherent contribution from three resonances, $X(6600)$, $X(6900)$, and $X(7100)$. The total event density is
\begin{equation}
F(m)
=
N_{\mathrm{NRSPS}}P_{\mathrm{NRSPS}}(m)
+N_{\mathrm{DPS}}P_{\mathrm{DPS}}(m)
+N_{\mathrm{BW}_0}P_{\mathrm{BW}_0}(m)
+N_{\mathrm{sig}}P_{\mathrm{sig}}(m),
\label{eq:total_model}
\end{equation}
where $N_i$ is the yield of component $i$, and each $P_i(m)$ is a PDF normalized over $[6,15]~\mathrm{GeV}$. All four component densities vanish for $m<m_{\mathrm{th}}$, where $m_{\mathrm{th}}=2m_{J/\psi}$ is the $J/\psi J/\psi$ production threshold~\cite{ParticleDataGroup:2026rpp}. The DPS shape is described by a fixed parametrization derived from simulated events~\cite{CMS:2023owd,CMS:2025fpt}. The threshold component $\mathrm{BW}_0$ is modeled by a relativistic $S$-wave Breit--Wigner with floating mass and width and does not interfere with the other components. The three signal resonances are described by relativistic $S$-wave Breit--Wigner amplitudes, combined coherently following the CMS three-resonance interference model~\cite{CMS:2023owd,CMS:2025fpt}:
\begin{equation}
P_{\mathrm{sig}}(m)
\propto
\left|
r_1\,e^{i\phi_1}\,\mathrm{BW}_1(m)+\mathrm{BW}_2(m)+r_3\,e^{i\phi_3}\,\mathrm{BW}_3(m)
\right|^2,
\label{eq:signal_model}
\end{equation}
where $\mathrm{BW}_1$, $\mathrm{BW}_2$, and $\mathrm{BW}_3$ correspond to $X(6600)$, $X(6900)$, and $X(7100)$, respectively. The $X(6900)$ amplitude is taken as the reference, with $r_2=1$ and $\phi_2=0$. The parameters $r_1$ and $r_3$ are dimensionless relative magnitudes, while $\phi_1$ and $\phi_3$ are relative phases in radians. Detector-resolution convolution and efficiency corrections are not applied in this study.

The NRSPS component is described by
\begin{equation}
f_{\mathrm{NRSPS}}(m)
=
(m-m_{\mathrm{th}})^{\alpha}
\left[
1
-
\left(
\frac{1}{(15-m_{\mathrm{th}})^2}-\frac{p_1}{10}
\right)
(15-m)^2
\right]
\exp\!\left[
-\frac{(m-m_{\mathrm{th}})^{p_3}}{2\,p_2^{p_3}}
\right],
\label{eq:sps}
\end{equation}
where the parameters $\alpha=0.80567$, $p_1=0.32647$, and $p_3=0.53968$ are fixed, while the dimensionless parameter $p_2$ is allowed to float within $[0,1]$. The normalized NRSPS density is given by $P_{\mathrm{NRSPS}}(m)\propto f_{\mathrm{NRSPS}}(m)$. As $p_2$ approaches zero, the distribution becomes increasingly concentrated near threshold.

\begin{figure}[htbp]
    \centering
    \includegraphics[width=0.5\linewidth]{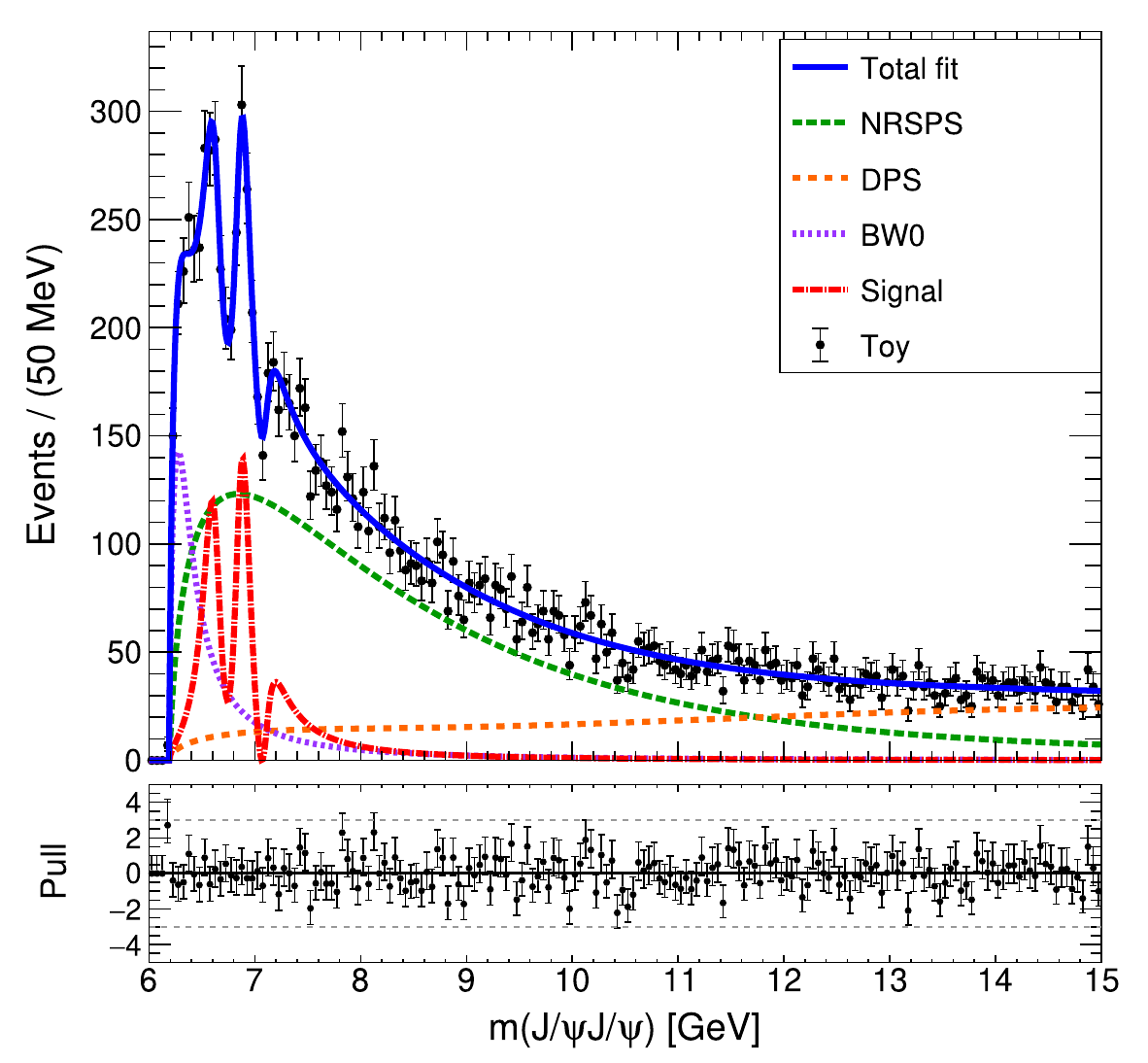}
    \caption{The $J/\psi J/\psi$ mass spectrum of the pseudo-dataset together with the result of a converged fit. The black points represent the pseudo-dataset, while the solid blue curve shows the total fit. The individual NRSPS, DPS, $\mathrm{BW}_0$, and coherent signal contributions are shown separately. The lower panel shows the pull distribution, defined as the difference between the data and the fit divided by the uncertainty in each bin.}
    \label{fig:toy_mass_spectrum}
\end{figure}

To examine numerical instabilities beyond a particular fit configuration of the CMS analysis~\cite{CMS:2023owd}, we retain the background parametrizations and their fixed parameters, while using different input values for the resonance masses, widths, interference parameters, component yields, and the NRSPS shape parameter $p_2$. A pseudo-dataset of 14,049 events, matching the event count in the CMS Run~2 dataset~\cite{CMS:2023owd}, is generated from this model. Figure~\ref{fig:toy_mass_spectrum} shows the resulting mass spectrum and the generation inputs are summarized in Tab.~\ref{tab:setup_params}. The same sample is used throughout the numerical studies.

The pseudo-dataset is fitted over the mass range of $6$--$15~\mathrm{GeV}$ using an extended unbinned maximum-likelihood fit. The negative log-likelihood (NLL) is defined as
\begin{equation}
\mathrm{NLL}
=
N_{\mathrm{tot}}
-
\sum_{j=1}^{N_{\mathrm{obs}}}
\ln
\left[
\sum_i N_iP_i(m_j)
\right],
\label{eq:extended_nll}
\end{equation}
where $N_{\mathrm{tot}}=\sum_iN_i$, and the index $i$ runs over the four model components. Each PDF is normalized numerically over the fitted mass range.
In the fit, a total of 17 parameters are allowed to vary: the four component yields, the masses and widths of $\mathrm{BW}_0$, $\mathrm{BW}_1$, $\mathrm{BW}_2$, and $\mathrm{BW}_3$, the relative interference parameters $r_1$, $r_3$, $\phi_1$, and $\phi_3$, and the NRSPS shape parameter $p_2$. Here $m_i$ and $\Gamma_i$ ($i=1,2,3$) denote the masses and widths of the three interfering resonances, while $m_{\mathrm{BW}_0}$ and $\Gamma_{\mathrm{BW}_0}$ describe the noninterfering threshold component. For the coherent signal, $r_2=1$ and $\phi_2=0$ are fixed. The remaining NRSPS shape parameters in Eq.~(\ref{eq:sps}) and the DPS shape parameters are also held fixed. A converged fit, consistent with the generation inputs, is overlaid in Fig.~\ref{fig:toy_mass_spectrum}. The allowed fit ranges and the fitted parameters are summarized in Tab.~\ref{tab:setup_params}.

\begin{table}[htbp]
\centering
\caption{
Generation inputs, allowed parameter ranges, and fitted values for the converged fit shown in Fig.~\ref{fig:toy_mass_spectrum}.
Masses and widths are given in GeV and relative phases in radians.
The parameters $r_2$ and $\phi_2$ are fixed, while the remaining 17 parameters are allowed to float in the fit.
The fitted values are broadly consistent with the generation inputs, with the largest deviation approximately twice the corresponding fit uncertainty.
}
\label{tab:setup_params}
\setlength{\tabcolsep}{3pt}
\begin{tabular}{llccc}
    \hline
    Parameter & Unit & Input & Fit range & Returned fit value \\
    \hline
    $N_{\mathrm{NRSPS}}$ & -- & 8153 & $[0,10^{5}]$ & $8071 \pm 335$ \\
    $N_{\mathrm{DPS}}$ & -- & 3645 & $[0,10^{5}]$ & $3168 \pm 235$ \\
    $N_{\mathrm{BW}_0}$ & -- & 1020 & $[0,10^{4}]$ & $1334 \pm 528$ \\
    $N_{\mathrm{sig}}$ & -- & 1231 & $[0,10^{4}]$ & $1476 \pm 363$ \\
    $m_{\mathrm{BW}_0}$ & GeV & 6.323 & $[6.20,6.50]$ & $6.339 \pm 0.045$ \\
    $\Gamma_{\mathrm{BW}_0}$ & GeV & 0.334 & $[0,1.50]$ & $0.413 \pm 0.199$ \\
    $m_1$ & GeV & 6.587 & $[6.40,6.80]$ & $6.635 \pm 0.025$ \\
    $\Gamma_1$ & GeV & 0.222 & $[0,1.50]$ & $0.215 \pm 0.084$ \\
    $r_1$ & -- & 1.217 & $[0,10^{3}]$ & $0.846 \pm 0.410$ \\
    $\phi_1$ & rad & $-0.334$ & $[-\pi,\pi]$ & $0.481 \pm 0.676$ \\
    $m_2$ & GeV & 6.871 & $[6.80,7.00]$ & $6.873 \pm 0.027$ \\
    $\Gamma_2$ & GeV & 0.134 & $[0,0.50]$ & $0.190 \pm 0.056$ \\
    $r_2$ & -- & 1 & fixed & 1 \\
    $\phi_2$ & rad & 0 & fixed & 0 \\
    $m_3$ & GeV & 7.134 & $[7.00,7.50]$ & $7.096 \pm 0.034$ \\
    $\Gamma_3$ & GeV & 0.097 & $[0,0.50]$ & $0.176 \pm 0.077$ \\
    $r_3$ & -- & 0.481 & $[0,10^{3}]$ & $0.653 \pm 0.260$ \\
    $\phi_3$ & rad & $-0.951$ & $[-\pi,\pi]$ & $-1.547 \pm 0.445$ \\
    $p_2$ & -- & 0.091 & $[0,1]$ & $0.107 \pm 0.010$ \\
    \hline
\end{tabular}
\end{table}

The likelihood fits are performed using RooFit and Minuit in ROOT~6.20.06~\cite{Verkerke:2003ir,James:1975dr,ROOT:website}. The NLL is minimized using the MIGRAD algorithm in Minuit, with RooIntegrator1D used to evaluate the numerical normalization integrals. MIGRAD is a variable-metric minimization algorithm that uses gradient information to iteratively approach the minimum of the NLL. The convergence of MIGRAD is primarily assessed using the estimated distance to minimum (EDM), which estimates the expected further decrease in the NLL. In the default RooFit configuration, the EDM is required to be below $10^{-3}$ for convergence. The integrator uses absolute and relative tolerances to determine when the estimated numerical error is sufficiently small to terminate integration. A more detailed explanation of the tolerance is given in Sec.~\ref{sec:discontinuities}. In the baseline configuration, both tolerances are set to $10^{-7}$, corresponding to the RooFit configuration used in this study. The integration method and tolerances can be configured through RooFit's \texttt{RooAbsReal::defaultIntegratorConfig()} interface. The HESSE procedure~\cite{James:1975dr} is subsequently used to estimate the Hessian matrix, whose elements are the second derivatives of the NLL with respect to the fit parameters at the fitted minimum, and to obtain the parameter covariance matrix and uncertainties from the inverse Hessian matrix.

\section{Observation of Anomalous NLL Behavior}
\label{sec:observation}

To ensure that the true global minimum of the NLL is found, the parameter space is explored by repeating the fit of the mass distribution 500 times on the same pseudo-dataset, with randomized initial parameter values drawn from the allowed fit ranges (Tab.~\ref{tab:setup_params}).
All fits use the numerical configuration of Sec.~\ref{sec:setup}, with both the absolute and relative integration tolerances set to $\varepsilon=10^{-7}$, meaning that integration stops when the estimated absolute or relative numerical error falls below this value.
For each fit, we track the minimum NLL ($\mathrm{minNLL}$), the EDM estimated by MIGRAD from the NLL gradient and curvature, the status codes of the minimization and subsequent covariance-matrix evaluation, and the covariance-matrix quality (covQual). A status code of 0 indicates successful minimization or covariance-matrix computation, while a code of 4 reflects failure. A covQual of 3 indicates an accurate covariance matrix, while 2 corresponds to a matrix modified to ensure positive definiteness. Lower values reflect an approximate covariance matrix, potentially containing only diagonal elements, a matrix that was not calculated, or one of unknown quality. A fit is regarded as reliably converged only if both the minimization and covariance-matrix evaluation return status codes of 0, the EDM is below $10^{-3}$, the covQual is 3, and no floating parameter lies at the boundary of its allowed range.

\begin{figure}[htbp]
    \centering
    \begin{tikzpicture}
        \node[inner sep=0] (fig)
        {\includegraphics[width=0.8\linewidth]
        {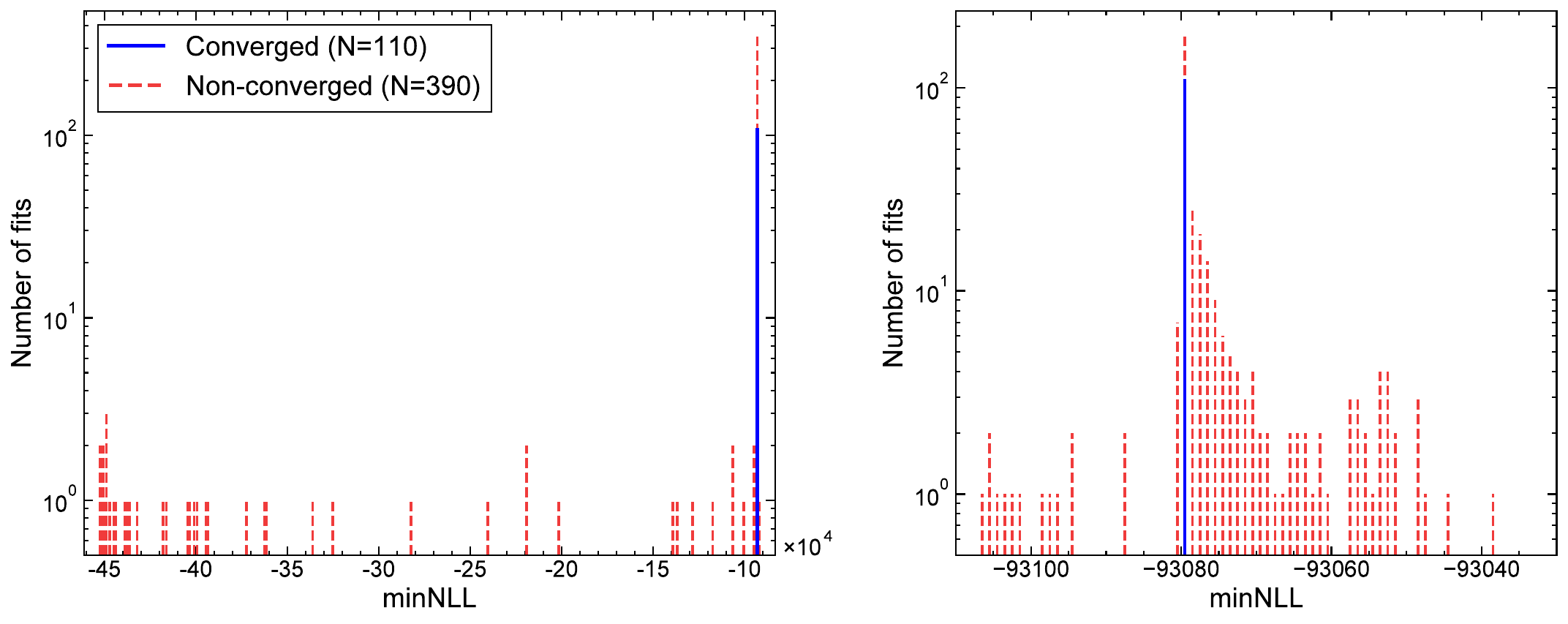}};
        \coordinate (xL) at ($(fig.south west)!0.007!(fig.south east)$);
        \coordinate (xR) at ($(fig.south west)!0.52!(fig.south east)$);
        \coordinate (yT) at ($(fig.south west)!0.96!(fig.north west)$);
        \node[anchor=north east, fill=white, inner sep=1pt]
        at (xL|-yT) {\textbf{(a)}};
        \node[anchor=north west, fill=white, inner sep=1pt]
        at (xR|-yT) {\textbf{(b)}};
    \end{tikzpicture}
    \caption{
    Distribution of $\mathrm{minNLL}$ for the 500 randomized fits, shown separately for fits that satisfy and that fail the convergence criteria.
    (a)~Full range.
    (b)~Zoom around the converged NLL region.
    }
    \label{fig:minnll_scan500}
\end{figure}

Among the 500 fits, 110 satisfy these criteria and converge to a common $\mathrm{minNLL}$ value of approximately $-93079.8$, which is used as a reference for comparison with the $\mathrm{minNLL}$ values obtained from the non-converged fits. Some non-converged fits, however, report substantially lower $\mathrm{minNLL}$ values. Figure~\ref{fig:minnll_scan500} shows the distribution of $\mathrm{minNLL}$ values obtained from the 500 fits with randomized initial parameters, distinguishing fits that satisfy the convergence criteria from those that do not.

Because each PDF is normalized by a numerically evaluated integral, integration errors can distort the NLL and affect the minimization. To test whether such errors contribute to the anomalous results, we select two sets of initial parameter values from the 500 fits performed at $\varepsilon=10^{-7}$: one associated with a converged fit reaching $\mathrm{minNLL}\approx-93079.8$, and the other with a non-converged fit reporting an anomalously low $\mathrm{minNLL}$. These are referred to as the reference and runaway starting points, respectively. To examine whether the anomalous fit behavior is sensitive to numerical integration accuracy, the minimization is repeated from each starting point with different absolute and relative integration tolerances of RooIntegrator1D, while all other settings remain unchanged. The results are summarized in Tab.~\ref{tab:precision_test}.

\begin{table}[htbp]
    \centering
    \caption{Fit results obtained from two sets of initial parameter values, referred to as the reference and runaway starting points. Both the absolute and relative RooIntegrator1D integration tolerances are set to the same value, denoted by $\varepsilon$. The first row ($\varepsilon=10^{-7}$) shows the original results of two fits selected from the 500 fits with randomized initial parameters: the fit from the reference starting point converged to $\mathrm{minNLL}\approx-93079.8$, whereas the fit from the runaway starting point failed to converge and returned an anomalously low $\mathrm{minNLL}$. Below the horizontal line, the fits are repeated using the same respective initial parameter values, with $\varepsilon$ varied from the looser $10^{-6}$ to the tighter $10^{-12}$. For each entry we report the fit status (0 for successful minimization and covariance-matrix evaluation, 4 for failure), the covariance-matrix quality (covQual; 3 for an accurate matrix, 2 for a matrix modified to ensure positive definiteness), the estimated distance to minimum (EDM) estimated by MIGRAD which estimates the expected further decrease in the NLL, and the minimized NLL ($\mathrm{minNLL}$). The two cases investigated in detail in Secs.~\ref{sec:discontinuities} and~\ref{sec:runaway}, corresponding to the reference starting point at $\varepsilon=10^{-6}$ and the runaway starting point at $\varepsilon=10^{-12}$, respectively, are highlighted in bold.}
    \label{tab:precision_test}
    \begin{tabular}{c|cccc|cccc}
        \hline
        & \multicolumn{4}{c|}{Reference starting point}
        & \multicolumn{4}{c}{Runaway starting point} \\
        \hline
        $\varepsilon$
        & status & covQual & EDM & minNLL
        & status & covQual & EDM & minNLL \\
        \hline
        $10^{-7}$ & 0 & 3 & $4.9\times10^{-5}$ & $-93079.8$
                  & 4 & 2 & $2.7\times10^{7}$ & $-449012.9$ \\
        \hline
        $\mathbf{10^{-6}}$ & $\mathbf{4}$ & $\mathbf{2}$ & $\mathbf{5.0\times10^{2}}$ & $\mathbf{-93081.1}$
                  & 4 & 2 & $3.1\times10^{2}$ & $-93081.1$ \\
        $10^{-8}$ & 0 & 3 & $5.1\times10^{-4}$ & $-93079.8$
                  & 0 & 3 & $2.3\times10^{-4}$ & $-93079.8$ \\
        $10^{-10}$ & 0 & 3 & $4.2\times10^{-4}$ & $-93079.8$
                   & 0 & 3 & $4.1\times10^{-4}$ & $-93079.8$ \\
        $10^{-12}$ & 0 & 3 & $1.1\times10^{-4}$ & $-93079.8$
                   & $\mathbf{4}$ & $\mathbf{3}$ & $\mathbf{4.7\times10^{14}}$ & $\mathbf{-119890.2}$ \\
        \hline
    \end{tabular}
\end{table}

At $\varepsilon=10^{-6}$, both starting points yield non-converged fits with $\mathrm{minNLL}\approx-93081.1$, about $1.3$~NLL units below the reference value of $-93079.8$ obtained from the 110 converged fits. At $\varepsilon=10^{-8}$ and $10^{-10}$, all fits return $\mathrm{minNLL}$ values consistent with the reference NLL and have fit status zero. At $\varepsilon=10^{-12}$, the runaway starting point produces a much more severe anomaly, with $\mathrm{minNLL}\approx-119890.2$ and an extremely large EDM. These results show that the reported NLL and the convergence behavior are highly sensitive to the numerical integration tolerance, even when the dataset, model, and starting parameters remain unchanged.
To understand the origin of these anomalies, the two cases highlighted in bold in Tab.~\ref{tab:precision_test} are investigated in the following sections, focusing first on the modest NLL reduction for the reference starting point at $\varepsilon=10^{-6}$ and then on the much more severe anomaly for the runaway starting point at $\varepsilon=10^{-12}$.

\section{Numerical Discontinuities in Likelihood Evaluation and Their Origin}
\label{sec:discontinuities}

\begin{figure}[htbp]
\centering

\noindent
\begin{minipage}[t]{0.48\linewidth}
    \centering
    \includegraphics[width=\linewidth]{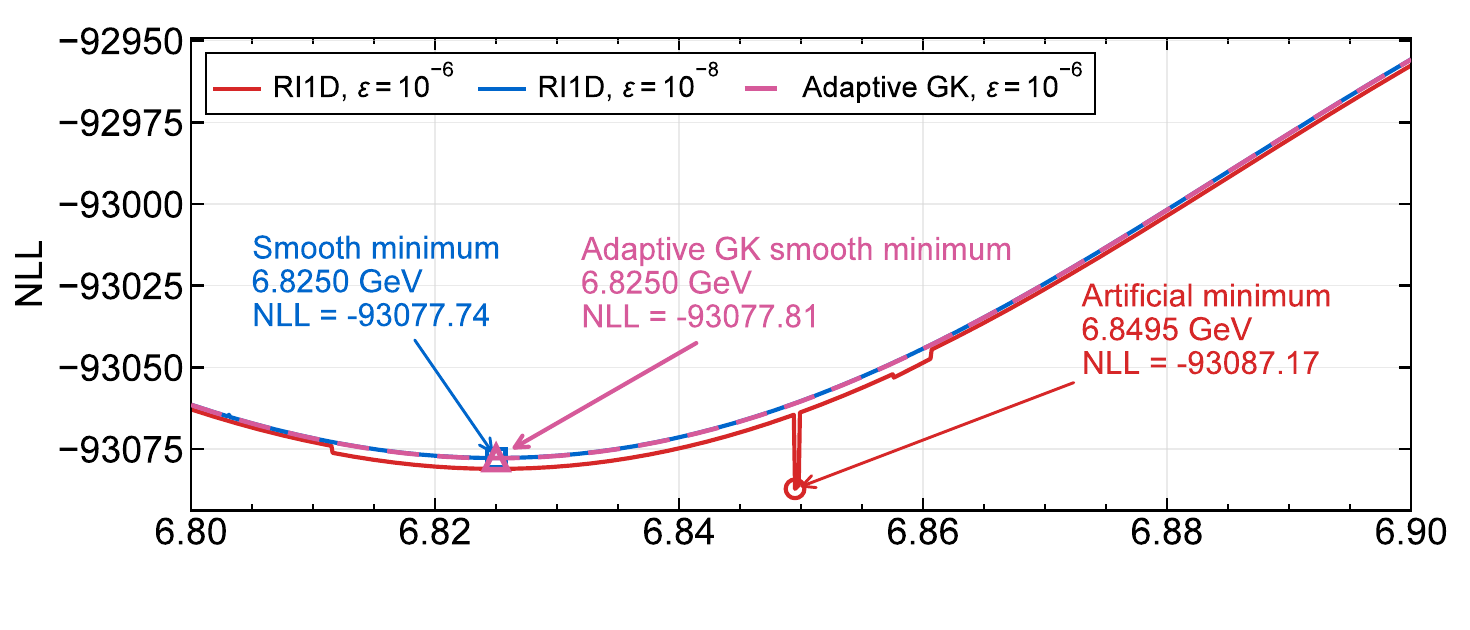}
    \par\vspace{1mm}
    (a)
\end{minipage}%
\hfill%
\begin{minipage}[t]{0.48\linewidth}
    \centering
    \includegraphics[width=\linewidth]{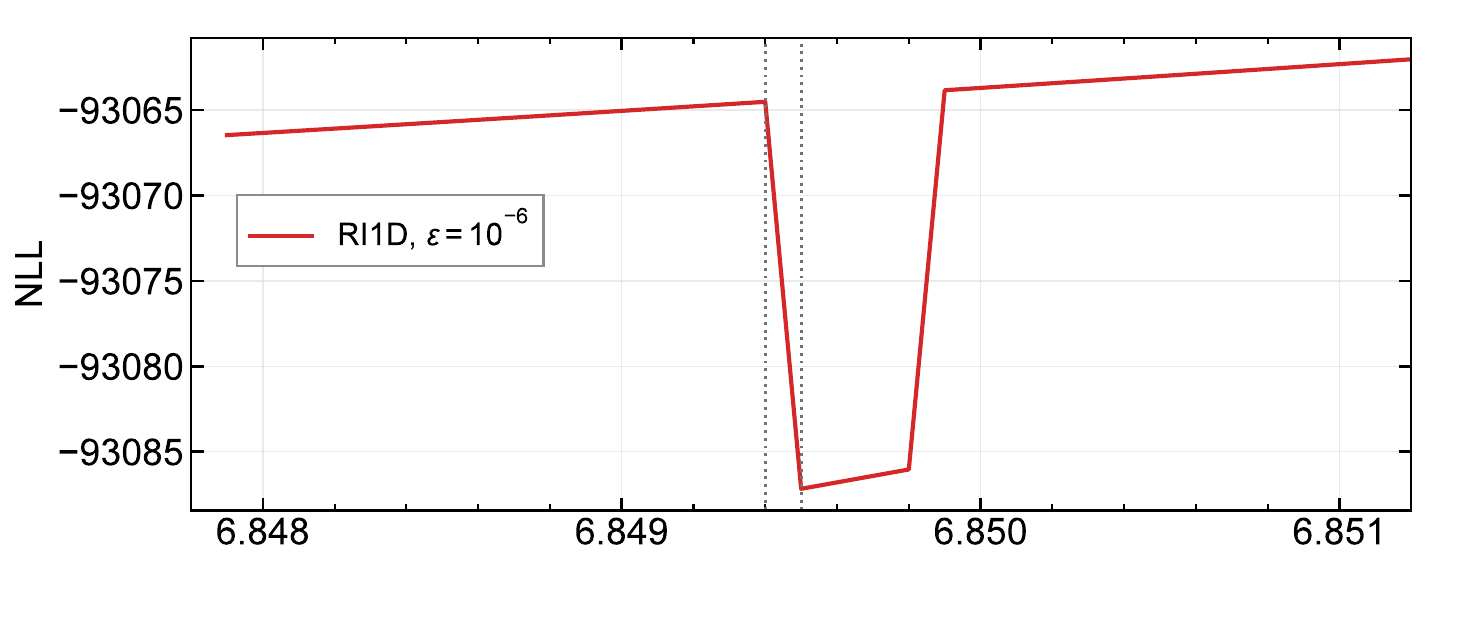}
    \par\vspace{1mm}
    (b)
\end{minipage}

\par\vspace{1mm}

\noindent
\begin{minipage}[t]{0.48\linewidth}
    \centering
    \includegraphics[width=\linewidth]{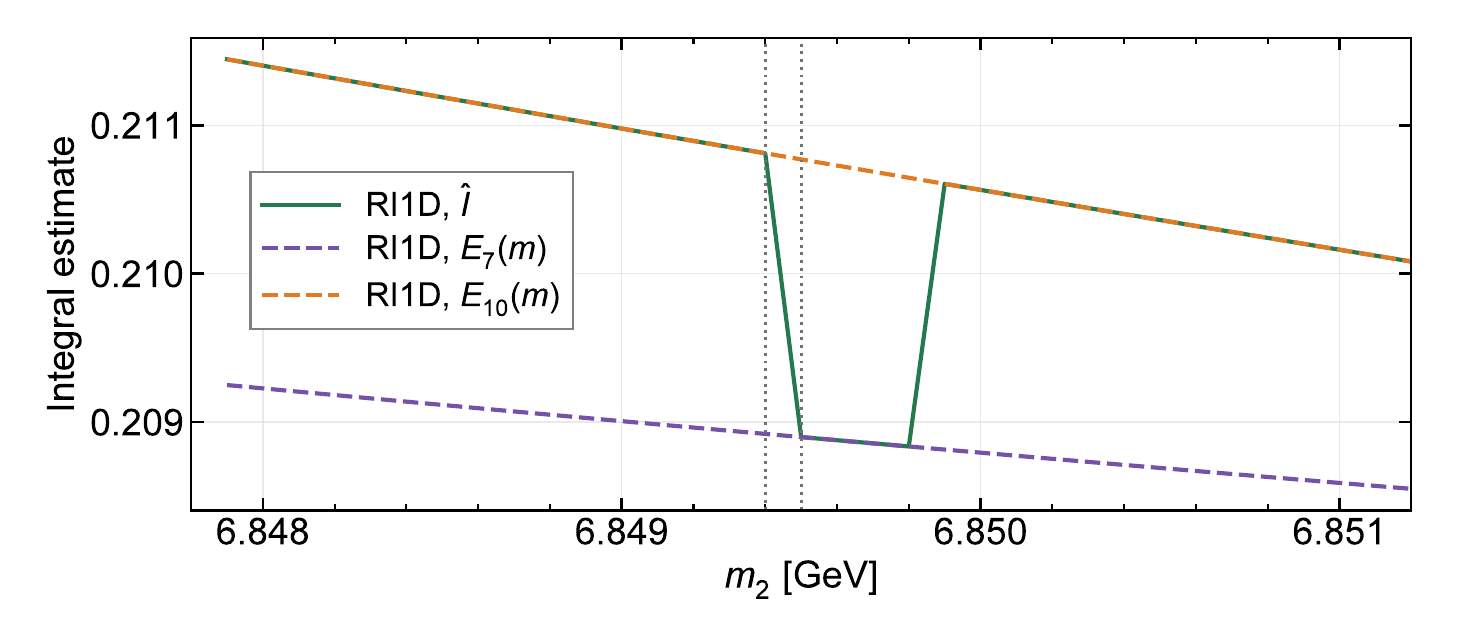}
    \par\vspace{1mm}
    (c)
\end{minipage}%
\hfill%
\begin{minipage}[t]{0.48\linewidth}
    \centering
    \includegraphics[width=\linewidth]{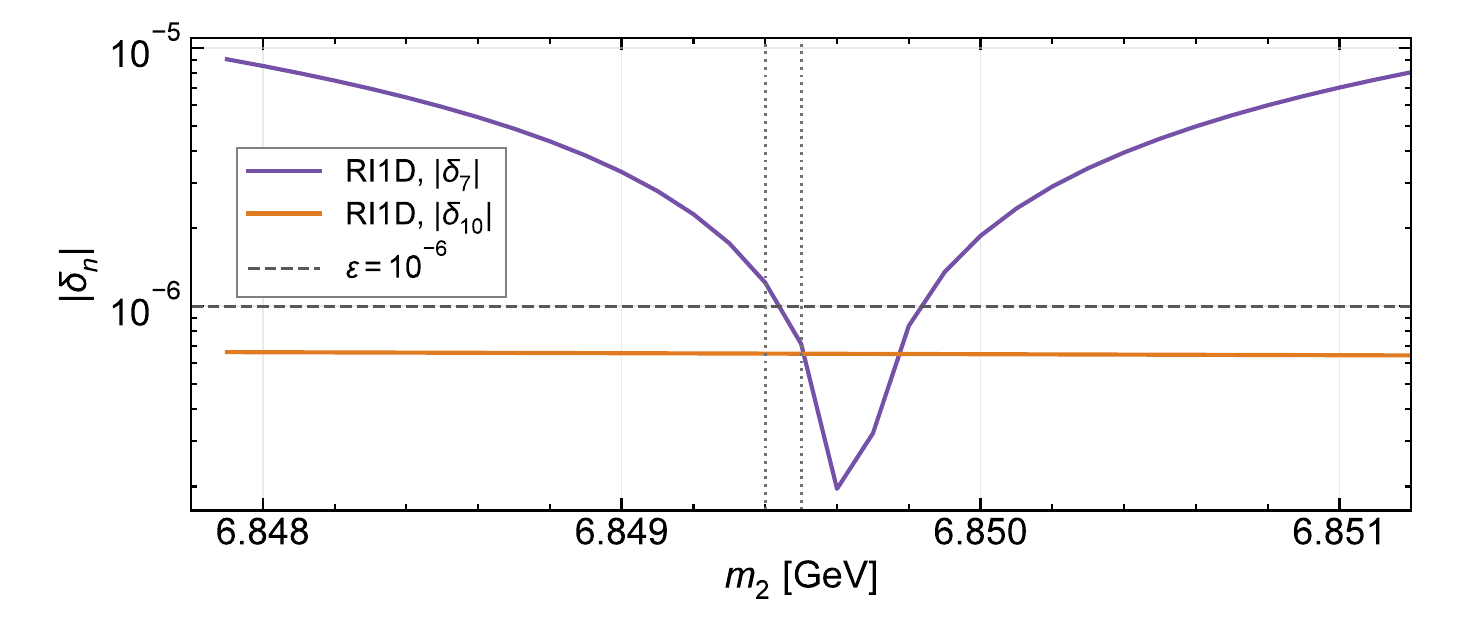}
    \par\vspace{1mm}
    (d)
\end{minipage}

\caption{
Numerical discontinuities and integrator stopping behavior in the $m_2$ scan for the reference starting point, with the other 16 fit parameters fixed at the values obtained from the fit using RooIntegrator1D (RI1D) at $\varepsilon=10^{-6}$.
(a)~NLL scans comparing RI1D with tolerances of $10^{-6}$ and $10^{-8}$, and adaptive Gauss--Kronrod (GK) integration with a tolerance of $10^{-6}$.
(b)~Enlarged view of the largest NLL discontinuity ($\Delta\mathrm{NLL}=-22.7$) obtained with RI1D at $\varepsilon=10^{-6}$.
(c)~Corresponding signal normalization integral returned by RI1D, $\hat{I}$, together with the integral estimates $E_7$ and $E_{10}$ obtained at refinement levels $n=7$ and $n=10$, respectively. RI1D evaluates the integral using the trapezoidal rule and progressively refines the integration grid by doubling the number of equal-width subintervals at each level, giving $2^{n-1}$ subintervals at level $n$. Here, $E_n$ denotes the integral estimate obtained through Neville extrapolation, and $n^{*}$ is the level at which the integrator stops.
(d)~Numerical error estimates $|\delta_7|$ and $|\delta_{10}|$ for the signal normalization integral at refinement levels $n=7$ and $n=10$, respectively, compared with the integration tolerance $\varepsilon=10^{-6}$.
Panels~(b)--(d) share the same $m_2$ range. The vertical dotted lines indicate the interval over which the integrator stopping level switches from $n^*=10$ to $n^*=7$ at a tolerance of $\varepsilon=10^{-6}$.
}
\label{fig:job11_discontinuity_stopping}
\end{figure}

To investigate the numerical behavior of the likelihood, a one-dimensional scan of the $X(6900)$ mass, $m_2$, is first examined as an illustrative example. For the non-converged fit obtained from the reference starting point using RooIntegrator1D at $\varepsilon=10^{-6}$ (Tab.~\ref{tab:precision_test}), all parameters are fixed to the values returned by that fit except $m_2$, which is scanned over $6.8$--$6.9~\mathrm{GeV}$ in steps of $10^{-4}~\mathrm{GeV}$. The scan exhibits pronounced discontinuities in the NLL (Fig.~\ref{fig:job11_discontinuity_stopping}(a)). To investigate their origin, we examine the numerical evaluation of the likelihood and its individual components in the vicinity of these discontinuities.

The largest NLL jump occurs between $m_2=6.8494$ and $6.8495~\mathrm{GeV}$, where $\Delta\mathrm{NLL}=-22.7$ (Fig.~\ref{fig:job11_discontinuity_stopping}(b)), compared with changes of approximately $0.1$--$0.4$ in the NLL over a $10^{-4}~\mathrm{GeV}$ step in $m_2$ away from the discontinuities.
Table~\ref{tab:app_reference_component_integrals} presents the NLL and the signal normalization integral $I_{\mathrm{sig}}$ around this discontinuity. As can be seen from the table, the jump is accompanied by an abrupt decrease of approximately $0.91\%$ in $I_{\mathrm{sig}}$, compared with changes of only about $0.01$--$0.02\%$ between other neighboring scan points. Although $I_{\mathrm{sig}}$ changes by less than $1\%$, its effect on the NLL is amplified by the presence of $2541$ signal events at this scan point. Varying only $I_{\mathrm{sig}}$ between the two adjacent scan points $m_2=6.8494$ and $6.8495~\mathrm{GeV}$, while holding the unnormalized densities fixed, yields $\Delta\mathrm{NLL}=-23.3$. The remaining contribution from the change in the unnormalized signal density is $+0.6$, giving a total NLL change of $-22.7$, in agreement with the observed jump. The normalization integrals of the DPS, NRSPS, and $\mathrm{BW}_0$ components remain unchanged, as expected because these components do not depend on $m_2$.

\begin{table}[htbp]
\centering
\caption{
NLL and signal normalization integral ($I_{\mathrm{sig}}$) evaluated in the $m_2$ scan around the largest NLL discontinuity, using RooIntegrator1D at $\varepsilon=10^{-6}$ with all other fit parameters held fixed. The differences $\Delta\mathrm{NLL}$ and $\Delta I_{\mathrm{sig}}$ are calculated relative to the preceding scan point, and $\Delta I_{\mathrm{sig}}/I_{\mathrm{sig}}$ is the corresponding relative change with respect to $I_{\mathrm{sig}}$ at that preceding point, reported as a percentage. The pairs of adjacent scan points spanning the two NLL discontinuities are highlighted in bold, together with the corresponding large changes in $\Delta\mathrm{NLL}$, $\Delta I_{\mathrm{sig}}$, and $\Delta I_{\mathrm{sig}}/I_{\mathrm{sig}}$.
}
\label{tab:app_reference_component_integrals}
\footnotesize
\setlength{\tabcolsep}{3pt}
\begin{tabular}{cccccc}
\hline
$m_2$ [GeV] & NLL & $I_{\rm sig}$ & $\Delta\mathrm{NLL}$ & $\Delta I_{\rm sig}$ & $\Delta I_{\rm sig}/I_{\rm sig}$ \\
\hline
6.8493 & $-93064.6$ & 0.21085 & --- & --- & --- \\
\textbf{6.8494} & \textbf{$-$93064.5} & \textbf{0.21081} & $0.1$ & $-0.00004$ & $-0.02\%$ \\
\textbf{6.8495} & \textbf{$-$93087.2} & \textbf{0.20890} & \textbf{$-$22.7} & \textbf{$-$0.00191} & \textbf{$-$0.91\%} \\
6.8496 & $-93086.8$ & 0.20888 & $0.4$ & $-0.00002$ & $-0.01\%$ \\
6.8497 & $-93086.4$ & 0.20886 & $0.4$ & $-0.00002$ & $-0.01\%$ \\
\textbf{6.8498} & \textbf{$-$93086.0} & \textbf{0.20883} & $0.4$ & $-0.00002$ & $-0.01\%$ \\
\textbf{6.8499} & \textbf{$-$93063.8} & \textbf{0.21061} & \textbf{$+$22.2} & \textbf{$+$0.00177} & \textbf{$+$0.85\%} \\
6.8500 & $-93063.7$ & 0.21056 & $0.1$ & $-0.00004$ & $-0.02\%$ \\
\hline
\end{tabular}
\end{table}

To understand why the numerical normalization changes abruptly as $m_2$ varies, we examine how RooIntegrator1D determines when to terminate numerical integration. To evaluate an integral $I=\int_a^b f(x)\,\mathrm{d}x$, RooIntegrator1D initially applies the trapezoidal rule over the full integration range $[a,b]$ using a single interval, giving $T_1=(b-a)[f(a)+f(b)]/2$. At each successive refinement level $n$, the number of equal-width subintervals is doubled, yielding a trapezoidal estimate $T_n$ based on $2^{n-1}$ subintervals~\cite{Romberg:1955}:
\[
T_n=\frac{T_{n-1}}{2}+h_n\sum_{k=1}^{2^{n-2}}f\bigl(a+(2k-1)h_n\bigr),\qquad
h_n=\frac{b-a}{2^{n-1}},\qquad
n\ge 2.
\]
As the grid spacing approaches zero, the trapezoidal estimates converge to the integral, $I=\lim_{n\to\infty}T_n$, and in practice this limit is estimated by extrapolating a finite number of estimates.
Once the first five estimates, $T_1,\ldots,T_5$, have been obtained, Neville's algorithm~\cite{Neville:1934} is used to extrapolate these estimates to the limit of vanishing grid spacing. At each level $n\ge 5$, the five most recent trapezoidal estimates are used to obtain an extrapolated estimate of the normalization integral $E_n$. The difference between the extrapolation using the four most recent estimates and that using all five, denoted $\delta_n$, serves as an internal error estimate.
Refinement continues until the estimated error $|\delta_n|$ falls below either the absolute tolerance $\varepsilon_{\mathrm{abs}}$ or the relative tolerance $\varepsilon_{\mathrm{rel}}|E_n|$,
\[
|\delta_n|
\leq
\max\left(\varepsilon_{\mathrm{abs}},
\varepsilon_{\mathrm{rel}}|E_n|\right),
\]
or until the maximum refinement level is reached. Once either tolerance is satisfied at level $n^*$, the integrator stops and returns the extrapolated estimate $E_{n^*}$. If the predefined maximum refinement level $n_{\max}$ is reached without convergence, the integrator reports non-convergence and returns the final trapezoidal estimate $T_{n_{\max}}$~\cite{ROOT:website}.
The RooFit default configuration sets the absolute and relative integration tolerances equal, with $\varepsilon_{\mathrm{abs}}=\varepsilon_{\mathrm{rel}}\equiv\varepsilon=10^{-7}$. We retain this convention and vary $\varepsilon$ to study the dependence on integration precision.
Since $\delta_n$ measures the last extrapolation correction rather than the deviation from the exact integral, a small $|\delta_n|$ does not necessarily guarantee an accurate result.

At the largest NLL discontinuity (Fig.~\ref{fig:job11_discontinuity_stopping}(b)), the stopping level changes from $n^{*}=10$ to $n^{*}=7$ between $m_2=6.8494$ and $6.8495~\mathrm{GeV}$. At the first point, the integration proceeds to level $n=10$ and returns the extrapolated estimate $E_{10}$. At the neighboring point, the internal error estimate $|\delta_7|$ falls below the stopping threshold of $10^{-6}$ (Fig.~\ref{fig:job11_discontinuity_stopping}(d)), causing the integrator to terminate at level $n=7$ and return $E_7$ instead. The switch between $E_{10}$ and $E_7$ introduces a discontinuity in the returned normalization integral, as shown in Fig.~\ref{fig:job11_discontinuity_stopping}(c). When $|\delta_7|$ rises above the stopping threshold again, the stopping level switches back to $n^{*}=10$, producing the second discontinuity near $m_2=6.8499~\mathrm{GeV}$.

\begin{figure}[htbp]
    \centering
    \begin{minipage}{0.42\linewidth}
        \centering
        \includegraphics[width=\linewidth]
        {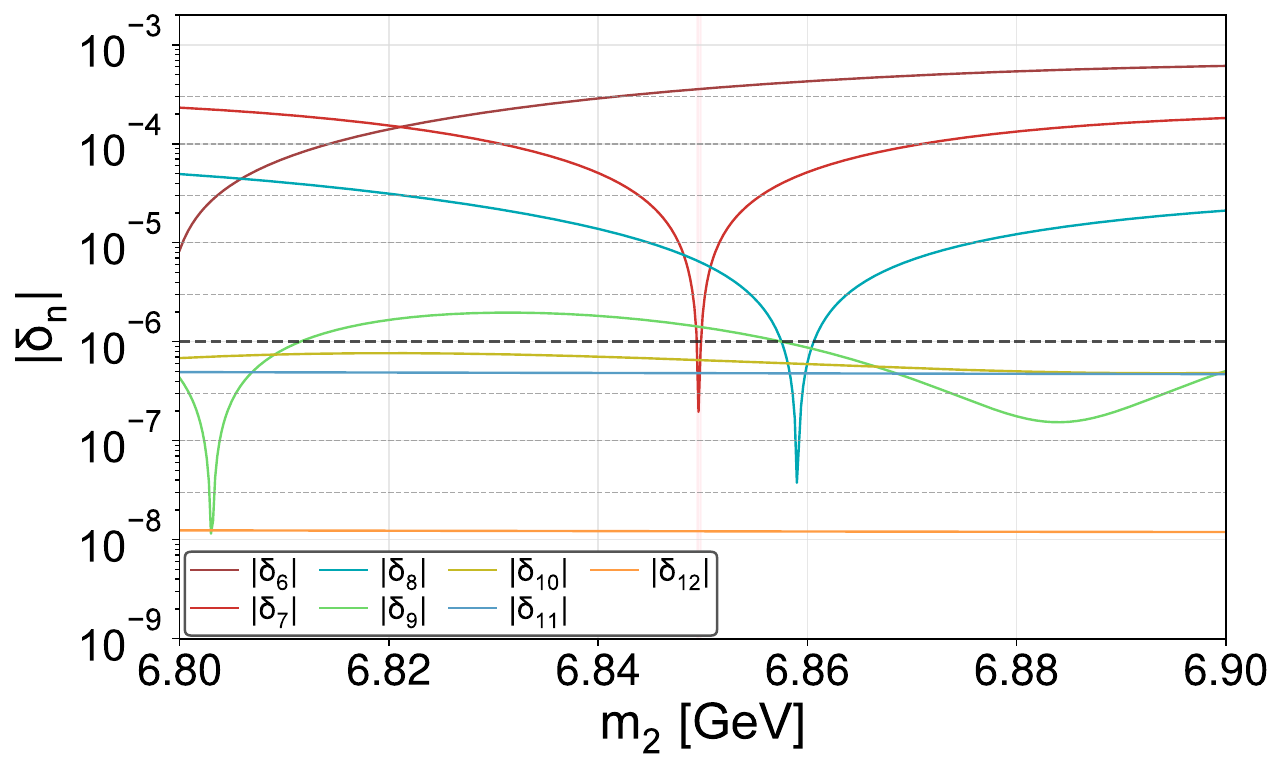}
        \par\vspace{1mm}
        (a)
    \end{minipage}
    \hfill
    \begin{minipage}{0.56\linewidth}
        \centering
        \begin{tikzpicture}
            \node[inner sep=0] (fig)
            {\includegraphics[width=\linewidth]
            {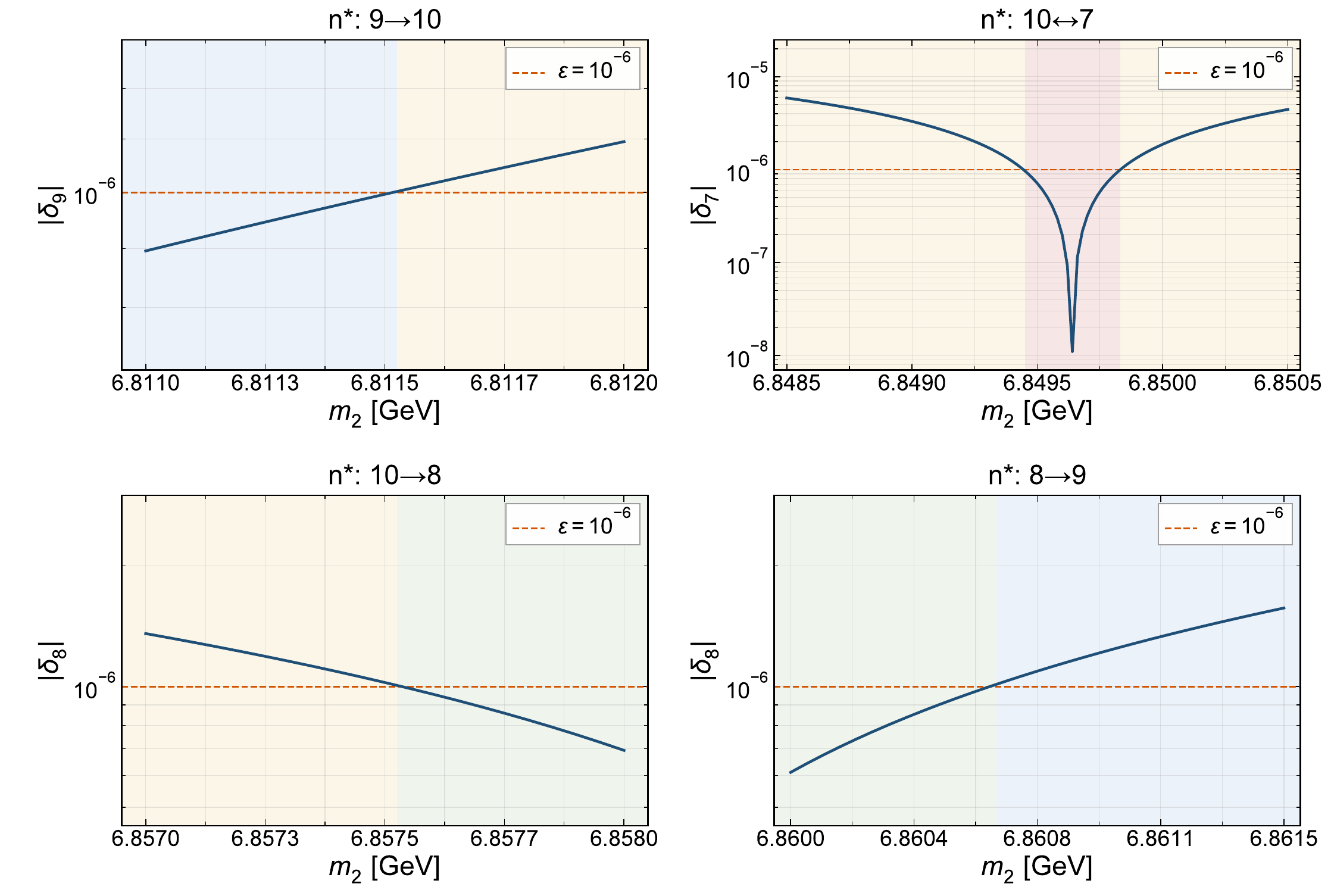}};
            \coordinate (xL) at ($(fig.south west)!0.02!(fig.south east)$);
            \coordinate (xR) at ($(fig.south west)!0.50!(fig.south east)$);
            \coordinate (yT) at ($(fig.south west)!0.98!(fig.north west)$);
            \coordinate (yB) at ($(fig.south west)!0.46!(fig.north west)$);
            \node[anchor=north west, fill=white, inner sep=1pt, yshift=3pt]
            at (xL|-yT) {\textbf{(i)}};
            \node[anchor=north west, fill=white, inner sep=1pt, yshift=3pt]
            at (xR|-yT) {\textbf{(ii)}};
            \node[anchor=north west, fill=white, inner sep=1pt, yshift=3pt]
            at (xL|-yB) {\textbf{(iii)}};
            \node[anchor=north west, fill=white, inner sep=1pt, yshift=3pt]
            at (xR|-yB) {\textbf{(iv)}};
        \end{tikzpicture}
        \par\vspace{1mm}
        (b)
    \end{minipage}
    \caption{
    Numerical error estimates and stopping-level transitions in the $m_2$ scan using RooIntegrator1D at $\varepsilon=10^{-6}$.
    (a)~Magnitudes of the Neville extrapolation corrections, $|\delta_n|$, for refinement levels $n=6$--$12$. The horizontal dashed line indicates the integration stopping threshold.
    (b)~Enlarged views of the error estimates near the stopping-level transitions associated with the NLL discontinuities in Fig.~\ref{fig:job11_discontinuity_stopping}(a). The solid curves show (i)~$|\delta_9|$, (ii)~$|\delta_7|$, (iii)~$|\delta_8|$, and (iv)~$|\delta_8|$, while the horizontal dashed lines indicate the stopping threshold. The shaded regions distinguish different stopping levels $n^*$, with transitions of (i)~$9\to10$, (ii)~$10\to7\to10$, (iii)~$10\to8$, and (iv)~$8\to9$. These transitions occur where the corresponding error estimates cross the stopping threshold.
    }
    \label{fig:app_reference_all_flips}
\end{figure}

Except for the discontinuity between $m_2=6.8494$ and $6.8495~\mathrm{GeV}$, there are four more discontinuities in the NLL for the $m_2$ scan, as shown in Fig.~\ref{fig:job11_discontinuity_stopping}(a) and summarized in Tab.~\ref{tab:app_reference_all_jumps}. All five NLL discontinuities coincide with a change in the refinement level $n^{*}$ at which RooIntegrator1D terminates. To examine these stopping-level changes, Fig.~\ref{fig:app_reference_all_flips}(a) shows the estimated integration errors $|\delta_n|$ at different refinement levels across the $m_2$ scan, while Fig.~\ref{fig:app_reference_all_flips}(b) provides enlarged views near the stopping-level transitions. At these transitions, an error estimate crosses the stopping threshold, changing the first refinement level that satisfies the stopping criterion. This causes the integrator to return a different normalization estimate, producing the corresponding NLL discontinuity.

\begin{table}[htbp]
\centering
\caption{
Changes in the NLL and signal normalization integral ($I_{\mathrm{sig}}$), together with the corresponding stopping-level transitions, across the five discontinuities identified in the $m_2$ scan in Fig.~\ref{fig:job11_discontinuity_stopping}(a), using RooIntegrator1D at $\varepsilon=10^{-6}$ with all other fit parameters held fixed. The $m_2$ transition column gives the two adjacent scan points spanning each discontinuity, while the $n^*$ transition indicates the corresponding change in the integration stopping level. The differences $\Delta\mathrm{NLL}$ and $\Delta I_{\mathrm{sig}}$ are calculated by subtracting the values at the first scan point from those at the second.
}
\label{tab:app_reference_all_jumps}
\footnotesize
\setlength{\tabcolsep}{3pt}
\begin{tabular}{cccc}
\hline
$m_2$ transition [GeV] & $n^*$ transition & $\Delta\mathrm{NLL}$ & $\Delta I_{\rm sig}$ \\
\hline
6.8115 $\rightarrow$ 6.8116 & $9\rightarrow10$ & $-2.2$ & $-2.75\times10^{-4}$ \\
6.8494 $\rightarrow$ 6.8495 & $10\rightarrow7$ & $-22.7$ & $-1.91\times10^{-3}$ \\
6.8498 $\rightarrow$ 6.8499 & $7\rightarrow10$ & $+22.2$ & $+1.77\times10^{-3}$ \\
6.8575 $\rightarrow$ 6.8576 & $10\rightarrow8$ & $-1.0$ & $-1.32\times10^{-4}$ \\
6.8606 $\rightarrow$ 6.8607 & $8\rightarrow9$ & $+2.8$ & $+1.87\times10^{-4}$ \\
\hline
\end{tabular}
\end{table}

\begin{figure}[htbp]
    \centering
    \begin{tikzpicture}
        \node[inner sep=0] (fig) {
            \includegraphics[width=0.95\linewidth]
            {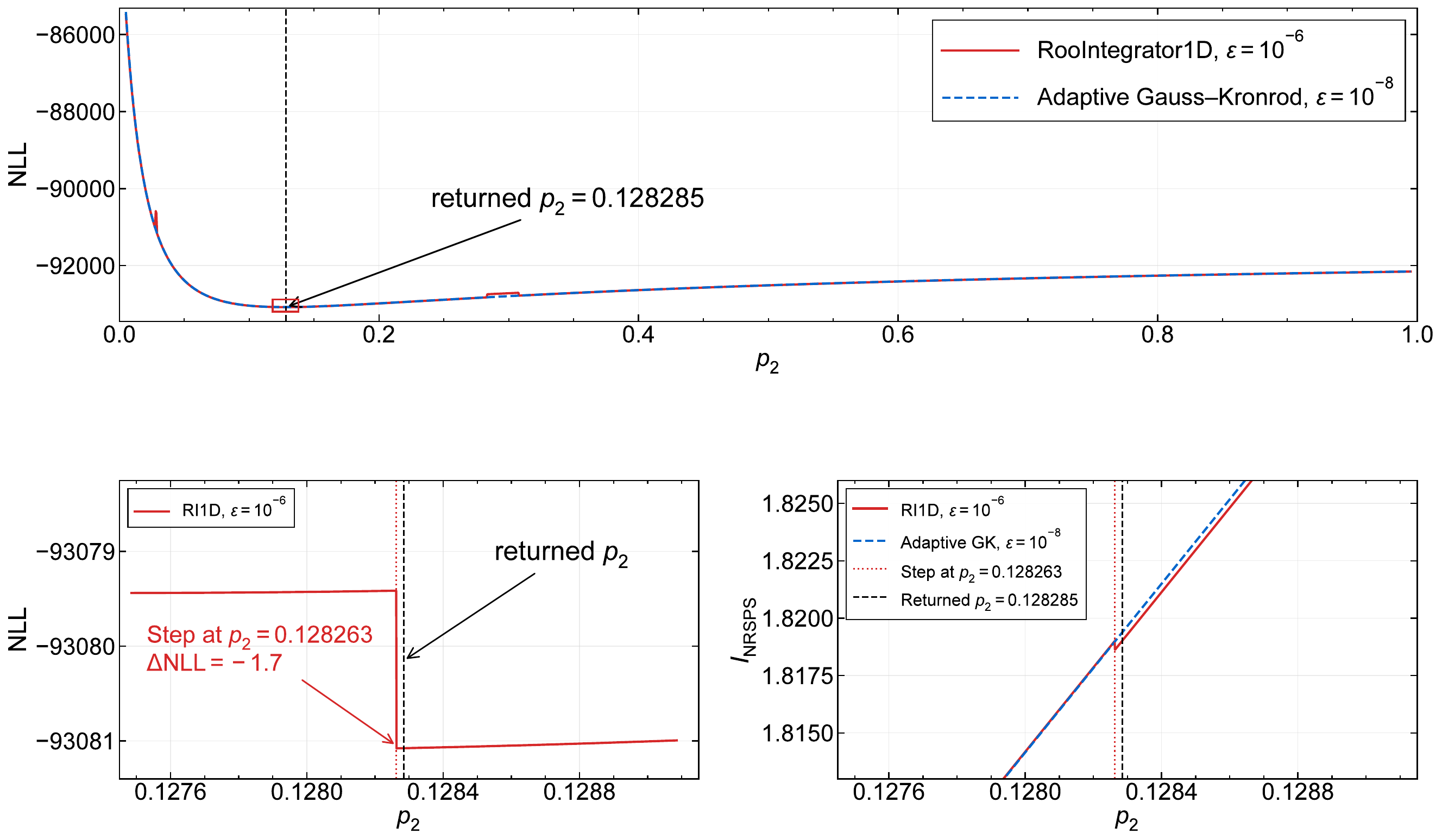}
        };

    \path (fig.south west) -- (fig.north west)
        coordinate[pos=0.53] (aheight);

       \node[anchor=center, fill=white, inner sep=1pt]
        at (aheight -| fig.center)
        {\textbf{(a)}};

        \node[anchor=north, inner sep=1pt]
        at ($(fig.south west)!0.25!(fig.south east)$)
        {\textbf{(b)}};

        \node[anchor=north, inner sep=1pt]
        at ($(fig.south west)!0.75!(fig.south east)$)
        {\textbf{(c)}};

    \end{tikzpicture}

    \caption{
    Numerical behavior of the likelihood as a function of the NRSPS shape parameter $p_2$, with the other 16 fit parameters fixed at the values returned by the non-converged fit using RooIntegrator1D (RI1D) at $\varepsilon=10^{-6}$.
    (a)~NLL as a function of $p_2$ over the full range, comparing RI1D at $\varepsilon=10^{-6}$ (red solid) with adaptive Gauss--Kronrod (GK) integration at $\varepsilon=10^{-8}$ (blue dashed).
    The red box indicates the region enlarged in panels~(b) and~(c).
    (b)~Enlarged NLL scan using RI1D at $\varepsilon=10^{-6}$, showing a numerical discontinuity of $\Delta\mathrm{NLL}=-1.7$ near the returned fit point.%
    (c)~Corresponding NRSPS normalization integrals evaluated using RI1D (red solid) and adaptive GK (blue dashed) with the same respective tolerances as in panel~(a).
    The red dotted and black dashed vertical lines mark the numerical discontinuity at $p_2=0.128263$ obtained with RI1D at $\varepsilon=10^{-6}$ and the corresponding returned fit value, respectively.%
    }
    \label{fig:job11_p2_roughness}
\end{figure}

Although the $m_2$ scan reveals how numerical integration can introduce artificial discontinuities in the NLL, the largest jump occurs approximately $24~\mathrm{MeV}$ away from the value $m_2=6.8252~\mathrm{GeV}$ at which the minimizer stopped, where the NLL is locally smooth along this parameter direction. The role of this discontinuity in the convergence failure is therefore not immediately clear. To investigate this further, we examine the NLL as a function of the other floating parameters while keeping the remaining parameters fixed at their returned values. Among these scans, a numerical step is observed along the NRSPS shape parameter $p_2$, close to its returned value (Figs.~\ref{fig:job11_p2_roughness}(a) and~\ref{fig:job11_p2_roughness}(b)). We therefore examine this step in detail to assess its role in the convergence failure.

The discontinuity occurs near $p_2=0.128263$, just below the returned value of $p_2=0.128285$, and is accompanied by a corresponding jump in the NRSPS normalization (Figs.~\ref{fig:job11_p2_roughness}(b) and~\ref{fig:job11_p2_roughness}(c)). Between two adjacent scan points, $p_2=0.128262$ and $0.128263$, the NLL decreases by approximately $1.7$, an unphysically large change for so fine a step in $p_2$, and the returned NRSPS normalization decreases by approximately $0.021\%$. This discontinuity arises from a change in the RooIntegrator1D stopping level from $n^{*}=11$ to $n^{*}=10$ between these two points.%
In the fit, MIGRAD performed approximately 2730 function evaluations, moving $p_2$ from its initial value of $0.144$ to $0.128285$, but failed to converge. At termination, the numerical first derivative associated with $p_2$ was approximately $-8\times10^{3}$, whereas those associated with the other 16 floating parameters were of order unity or smaller. This derivative is consistent with a finite-difference evaluation crossing the NLL discontinuity of approximately 1.7 and is much larger in magnitude than the local slope of the otherwise smooth NLL curve. MIGRAD failed to find a further decrease in the evaluated NLL and returned this point with an EDM of $5.0\times10^{2}$. The subsequent HESSE calculation failed to obtain a positive-definite error matrix. The large EDM and the absence of an accurate covariance matrix both fall short of the convergence criteria defined in Sec.~\ref{sec:observation}, indicating that the fit had not reached a trustworthy minimum.%

To investigate the role of the numerical discontinuity in the convergence failure, additional minimization tests were performed. Starting from the returned fit point, both a fit with only $p_2$ floating and a full 17-parameter fit failed to move away from that point using RooIntegrator1D at $\varepsilon=10^{-6}$. In contrast, repeating the full fit from the original initial values with the RooIntegrator1D refinement level fixed at $n=11$ restored convergence, with successful completion of the minimization and covariance-matrix evaluation ($\mathrm{status}=0$), an accurate covariance matrix ($\mathrm{covQual}=3$), and an estimated further decrease in the NLL below the $10^{-3}$ threshold ($\mathrm{EDM}=7.3\times10^{-5}$).%
These results indicate that the integration stopping-level switching was responsible for the observed convergence failure.

To mitigate these numerical discontinuities in the NLL, one approach is to tighten the integration tolerance $\varepsilon$. Alternatively, the integration can be improved by using a different numerical integration method. Since the accuracy of an integral depends on how the integration domain is sampled, adaptive Gauss--Kronrod quadrature~\cite{Galassi:2019czg}, for example, uses two numerical approximations of the same integral to estimate local integration errors and selectively refine subintervals, in contrast to the uniform refinement used by RooIntegrator1D.
Both approaches yield smooth NLL curves in the $m_2$ scan, as shown in Fig.~\ref{fig:job11_discontinuity_stopping}(a). The discontinuity near the returned $p_2$ value is also absent with adaptive Gauss--Kronrod integration at $\varepsilon=10^{-8}$ (Fig.~\ref{fig:job11_p2_roughness}). To assess the impact on minimization, the full 17-parameter fit is repeated from the same initial values using adaptive Gauss--Kronrod integration at $\varepsilon=10^{-8}$. The fit converges with $\mathrm{status}=0$, $\mathrm{covQual}=3$, and $\mathrm{EDM}=1.6\times10^{-4}$,%
reaching $\mathrm{minNLL}=-93079.8$, consistent with the reference value obtained in Sec.~\ref{sec:observation}. The recovery of convergence supports the interpretation that numerical discontinuities in the PDF normalization, and the resulting artificially lower NLL, contributed to the original minimization failure.

\section{Severe Numerical Integration Errors from Unresolved Narrow Structures}
\label{sec:runaway}

\subsection{Mechanism of Normalization Underestimation in the Presence of Narrow Structures}

As shown in Tab.~\ref{tab:precision_test}, repeating the fit from the same runaway starting point with progressively tighter integration tolerances leads to markedly different results. At $\varepsilon=10^{-8}$ and $10^{-10}$, the fits return $\mathrm{minNLL}$ values near $-93079.8$, consistent with the reference solutions. However, further tightening the tolerance to $\varepsilon=10^{-12}$ produces a non-converged fit with an anomalously low $\mathrm{minNLL}$ of $-119890.2$. In this fit, the NRSPS shape parameter $p_2$ decreases to $4.33\times10^{-6}$, much smaller than the input value of $0.091$, while the NRSPS yield increases to approximately $9.0\times10^{4}$, which is even larger than the total input yield of $14049$. These extreme values are clearly not a physically preferred description of the spectrum.

\begin{figure}[htbp]
    \centering
    \begin{minipage}{0.48\linewidth}
        \centering
        \includegraphics[width=\linewidth]
        {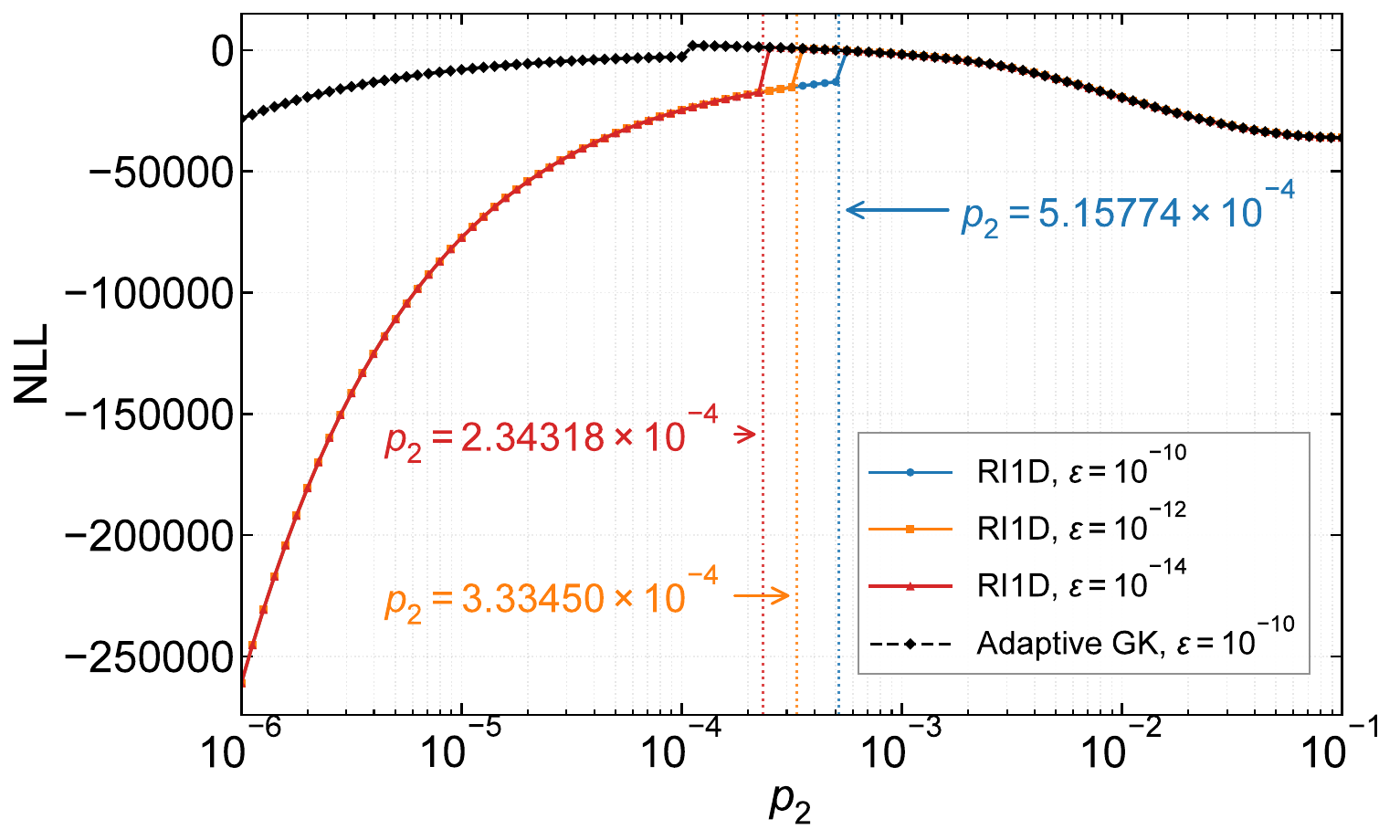}
        \par\vspace{1mm}
        (a)
    \end{minipage}
    \hfill
    \begin{minipage}{0.48\linewidth}
        \centering
        \includegraphics[width=\linewidth]
        {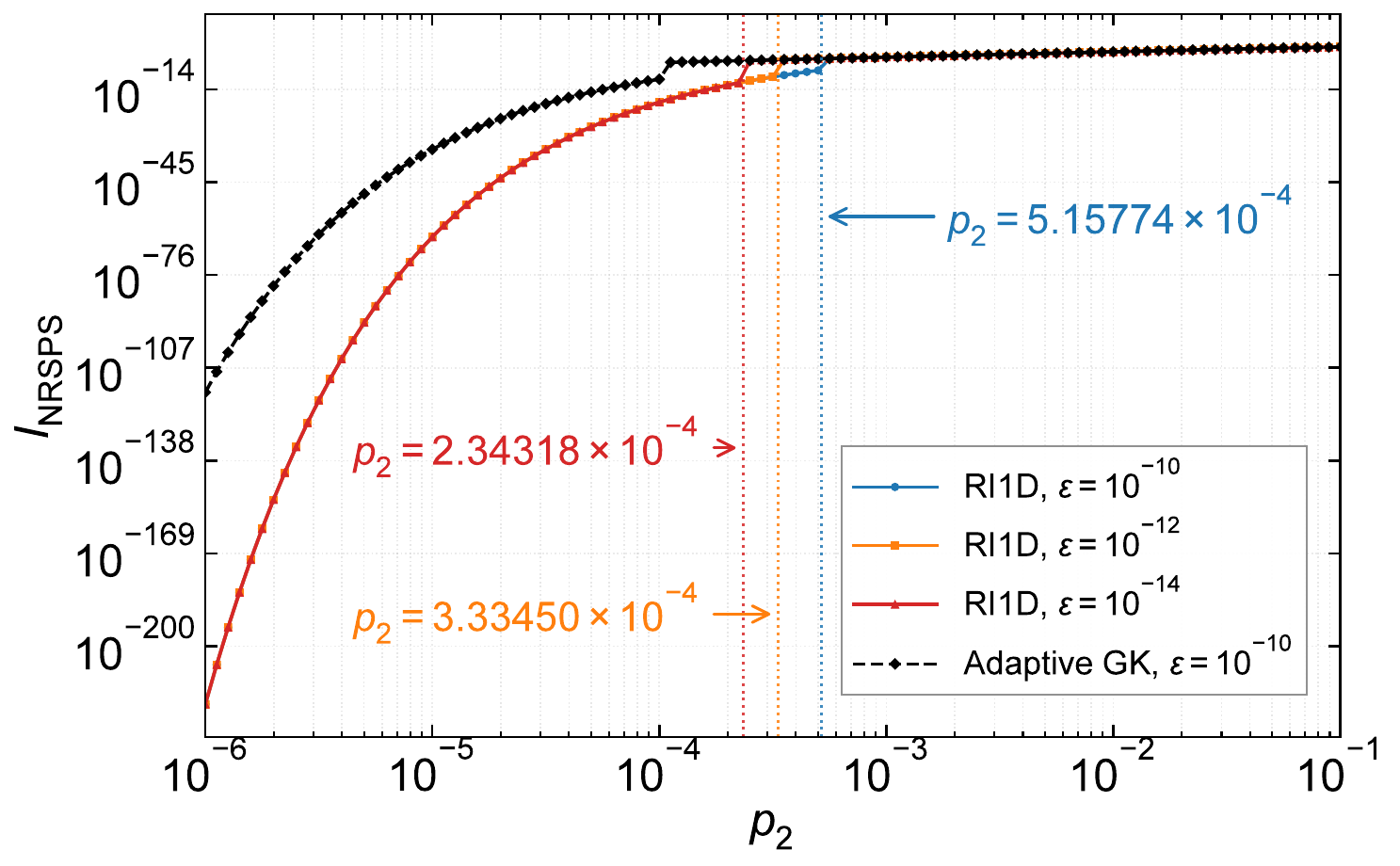}
        \par\vspace{1mm}
        (b)
    \end{minipage}

    \vspace{2mm}

    \begin{minipage}{0.48\linewidth}
        \centering
        \includegraphics[width=\linewidth]
        {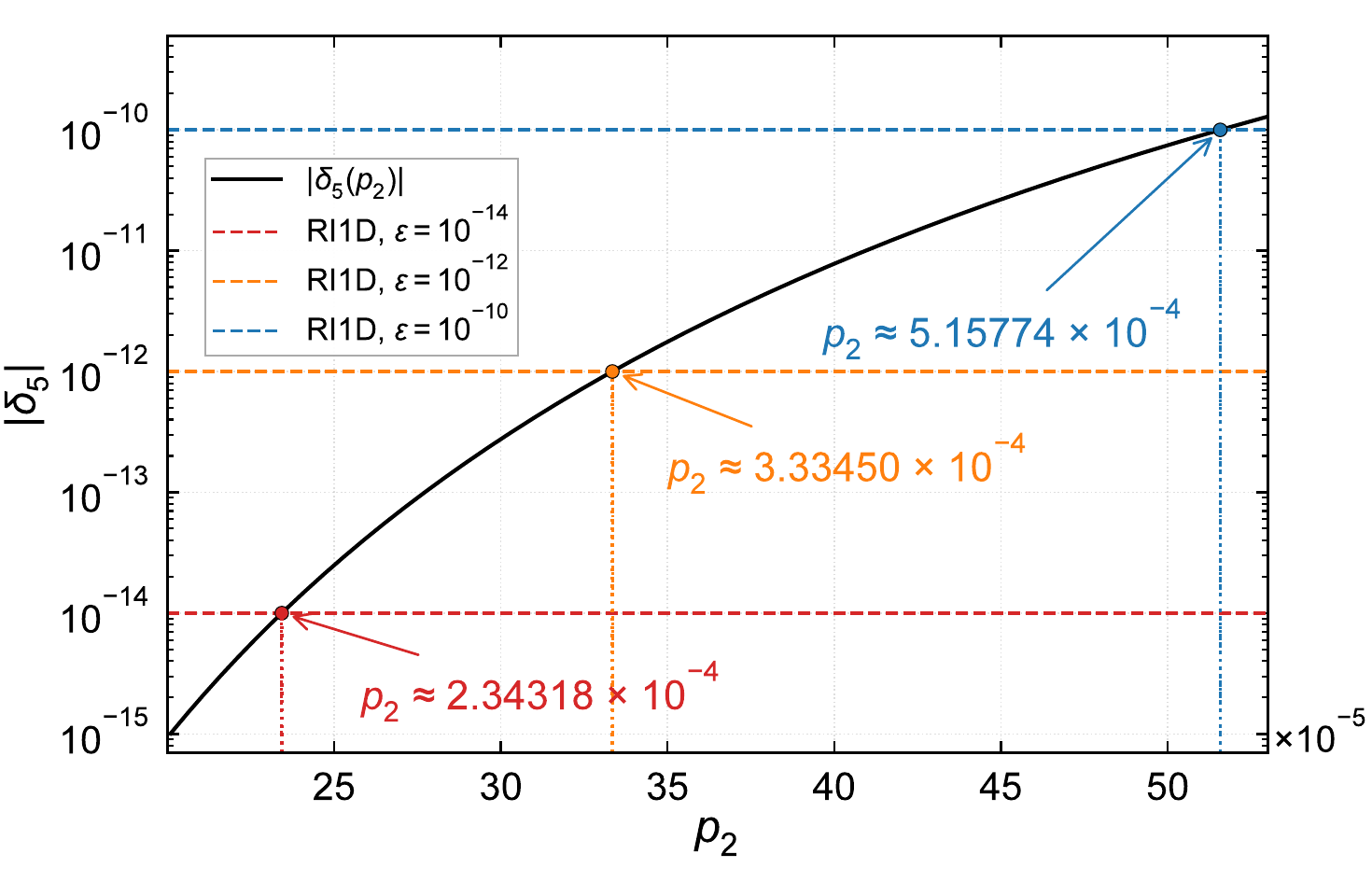}
        \par\vspace{1mm}
        (c)
    \end{minipage}
    \hfill
    \begin{minipage}{0.48\linewidth}
        \centering
        \includegraphics[width=\linewidth]
        {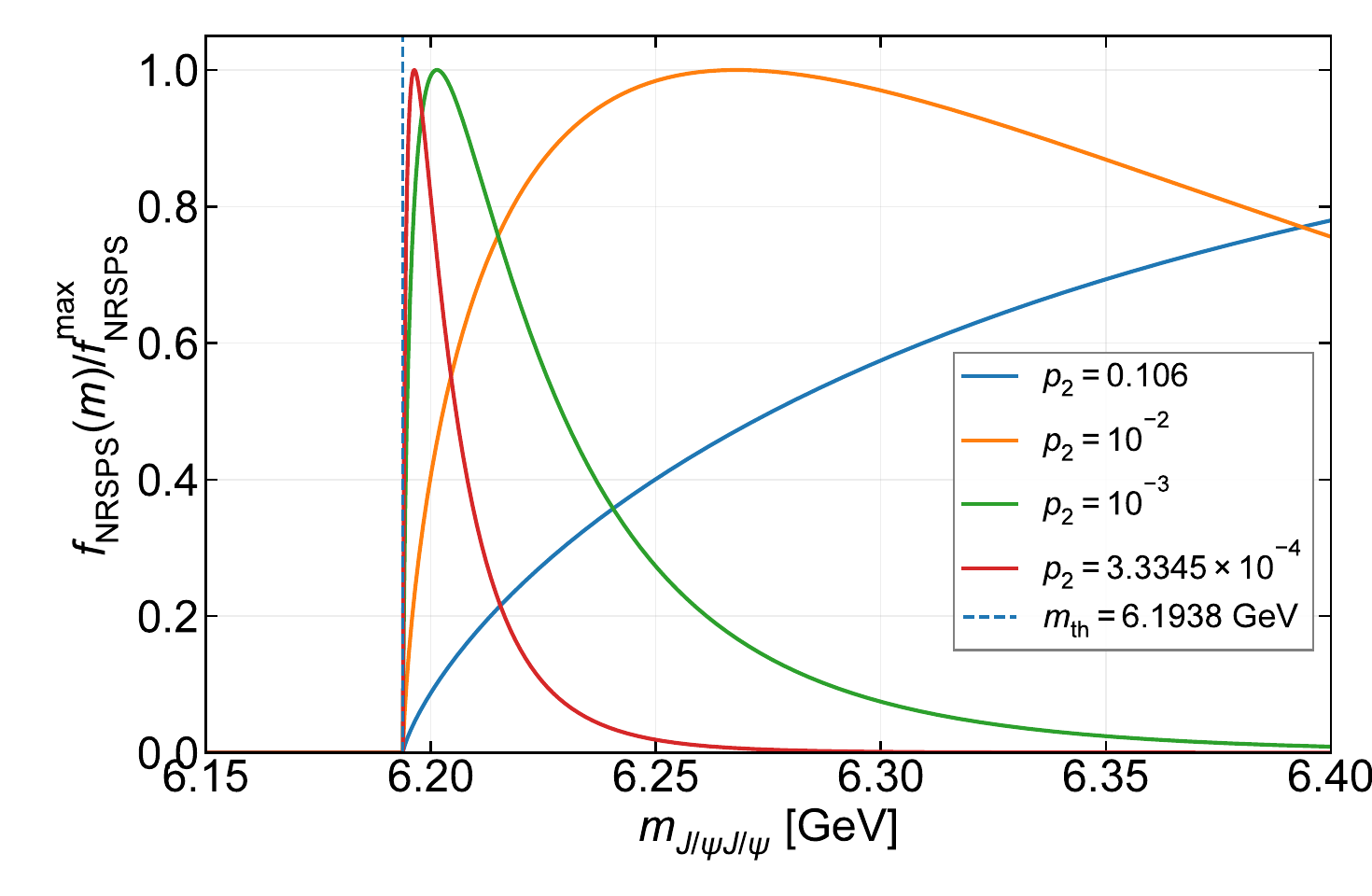}
        \par\vspace{1mm}
        (d)
    \end{minipage}
    \caption{
    Numerical behavior of the NRSPS normalization and likelihood as a function of $p_2$.
    (a)~NLL as a function of $p_2$, with the other 16 fit parameters fixed at the values returned by the runaway fit at $\varepsilon=10^{-12}$. Results are shown for RooIntegrator1D (RI1D) at $\varepsilon=10^{-10}$, $10^{-12}$, and $10^{-14}$, and adaptive Gauss--Kronrod (GK) integration at $\varepsilon=10^{-10}$.
    (b)~Corresponding NRSPS normalization integral $I_{\mathrm{NRSPS}}$ evaluated with the same integration settings. The vertical dotted lines in (a) and (b) mark the RI1D stopping transitions at $\varepsilon=10^{-14}$, $10^{-12}$, and $10^{-10}$, with the corresponding $p_2$ values annotated.
    (c)~Estimated integration error $|\delta_5|$ at refinement level $n=5$ as a function of $p_2$. The intersections of the $|\delta_5|$ curve with the horizontal dashed lines mark the stopping-transition positions for the three integration tolerances.
    (d)~NRSPS mass distributions for different $p_2$ values, each normalized to its maximum. The vertical dashed line indicates the $J/\psi J/\psi$ threshold at $m_{\mathrm{th}}=6.1938~\mathrm{GeV}$. The distributions become increasingly narrow near threshold as $p_2$ decreases.
    }
    \label{fig:runaway_overview}
\end{figure}

To investigate this behavior, we examine the NLL as a function of the floating parameters while keeping the remaining parameters fixed at the values returned by the runaway fit. Unlike the other parameter scans, the $p_2$ scan reveals a pronounced decrease in the NLL toward very small $p_2$ values (Fig.~\ref{fig:runaway_overview}(a)), accompanied by a strong suppression of the NRSPS normalization integral (Fig.~\ref{fig:runaway_overview}(b)). Both RooIntegrator1D and adaptive Gauss--Kronrod integration exhibit this behavior, although the suppression is substantially more severe with RooIntegrator1D. Discontinuities are also observed in the NLL and normalization curves. For RooIntegrator1D, the integrator first checks whether the estimated integration error satisfies the prescribed tolerance at refinement level $n=5$, using $|\delta_5|$ obtained from the extrapolation of successive trapezoidal approximations. As $p_2$ varies, $|\delta_5|$ crosses the integration tolerance (Fig.~\ref{fig:runaway_overview}(c)). On one side of the crossing, the integrator terminates at $n=5$, while on the other side, it continues to higher refinement levels. The resulting difference in the returned normalization estimates produces the observed discontinuities in the normalization and NLL. Similar discontinuities in adaptive Gauss--Kronrod integration also arise from changes in the integration stopping decision. These stopping transitions explain the observed discontinuities, but not why the NRSPS normalization becomes extremely small at low $p_2$.

The severe underestimation of the NRSPS normalization originates from the increasingly narrow distribution near the $J/\psi J/\psi$ threshold. As shown in Fig.~\ref{fig:runaway_overview}(d), decreasing $p_2$ concentrates the distribution into a progressively smaller mass interval above the threshold. At refinement level $n=5$, RooIntegrator1D divides the integration range of $6$--$15~\mathrm{GeV}$ into 16 equal subintervals. The first sampling point above the $J/\psi J/\psi$ threshold is therefore at $6.5625~\mathrm{GeV}$, approximately $369~\mathrm{MeV}$ above threshold, whereas the NRSPS peak lies only $2.5~\mathrm{MeV}$ above threshold at $p_2\simeq3.33\times10^{-4}$.%
The integration grid therefore misses the region that provides most of the integral and underestimates the true integral by many orders of magnitude. Since the estimated integration errors are calculated from the sampled function values, the errors may remain small even when the integration grid fails to resolve the dominant contribution, causing the integrator to terminate prematurely. As a result, tightening the integration tolerance does not necessarily prevent this failure. The same limitation can affect other numerical integration methods, including adaptive Gauss--Kronrod quadrature, whose error estimates also depend on sampled function values.

We further investigate how the minimizer moves from its initial value of $p_2=0.123$, bypasses the expected minimum near $p_2\simeq0.107$, and eventually enters the anomalous small-$p_2$ region.%
As shown in Tab.~\ref{tab:precision_test}, the fit from the same runaway starting point first fails to converge at looser tolerances. After the integration tolerance is tightened, the fits at $\varepsilon=10^{-8}$ and $10^{-10}$ return $\mathrm{minNLL}$ values near $-93079.8$, consistent with the reference solution, whereas the fit at $\varepsilon=10^{-12}$ returns a much lower value of $-119890.2$.
To understand why the two integration tolerances lead to different fit outcomes, we examine how MIGRAD~\cite{James:1975dr} explores the parameter space. During minimization, MIGRAD uses changes in the NLL to determine the direction in which the parameters should move and then evaluates the NLL at different distances along that direction to find a parameter point with a lower NLL.
To keep $p_2$ within its specified range of $[0,1]$, Minuit employs an internal parameter $u$ for minimization, which is mapped to $p_2$ through $p_2=(1+\sin u)/2$, following the Minuit standard transformation for parameters with finite lower and upper bounds. Because MIGRAD performs a line search in the internal parameter space, the nonlinear and periodic sine transformation allows a finite step in $u$ to reach a very small $p_2$ without evaluating the NLL at intermediate $p_2$ values between its current and new values, including intermediate values near the $p_2$ value of the normally converged fit.

For the fits from the same runaway starting point at $\varepsilon=10^{-10}$ and $\varepsilon=10^{-12}$, with only the former converging, the change in integration tolerance produces small differences in the evaluated NLL, typically of order $10^{-3}$ when the parameter values are nearly identical. These differences can affect the estimated gradients and consequently alter the search directions and the subsequent parameter points evaluated by MIGRAD. Although these two fits reach nearly identical parameter points, with $p_2\simeq0.123$, at the start of a line search, their subsequent evaluations probe different points in the 17-dimensional parameter space, with the trial values of $p_2$ differing by a factor of approximately 2.7 and the other fit parameters also taking different values. At $\varepsilon=10^{-12}$, severe underestimation of the NRSPS normalization at the trial point produces an artificially low NLL of approximately $-119890$, which is accepted as an improvement because the underestimated normalization inflates the NRSPS PDF contribution to the likelihood. In contrast, at $\varepsilon=10^{-10}$, the trial at a different small-$p_2$ point gives an NLL of approximately $-88994$, about 4055 higher than the lowest NLL previously found during that fit, and is therefore rejected. The minimizer subsequently returns to the vicinity of the normally converged region and eventually converges to the reference solution at $p_2\simeq0.107$. The different outcomes of these two fits show that changes in the integration tolerance can introduce small differences in the evaluated NLL, which in turn affect the parameter-space search performed by MIGRAD.

\subsection{Improving Numerical Integration of Narrow Structures through Variable Transformations}
\label{sec:solution}

The numerical failure identified above arises because the integration grid does not adequately resolve the narrow NRSPS contribution near threshold. To improve the sampling of that narrow region, we apply a M\"obius variable transformation~\cite{Homeier1990} to concentrate numerical sampling points near this region, without changing the NRSPS function or the likelihood definition.

\begin{figure}[htbp]
    \centering
    \begin{minipage}{0.32\linewidth}
        \centering
        \includegraphics[width=\linewidth]
        {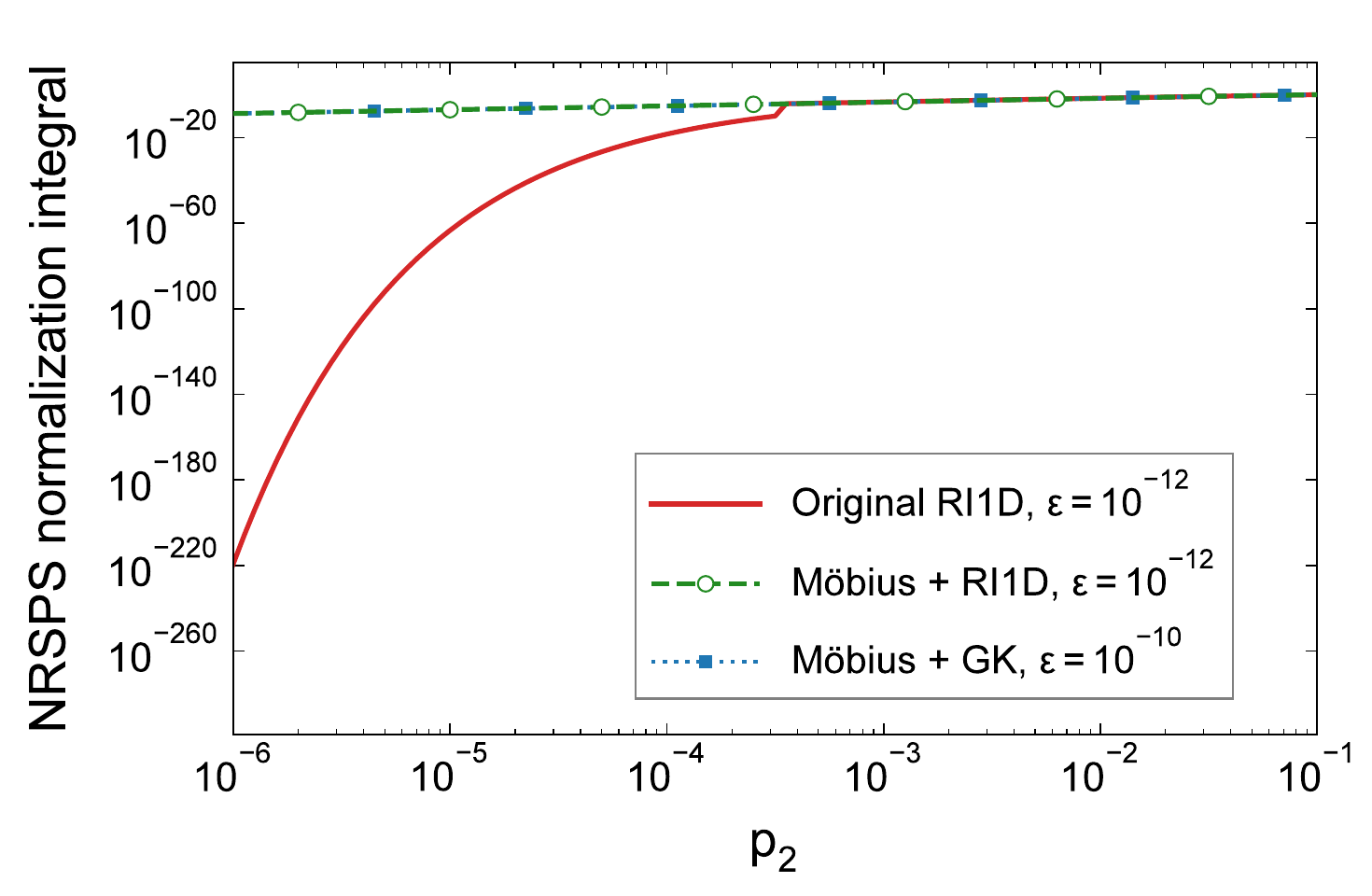}
        \par\vspace{1mm}
        (a)

    \end{minipage}
    \hfill
    \begin{minipage}{0.32\linewidth}
        \centering
        \includegraphics[width=\linewidth]
        {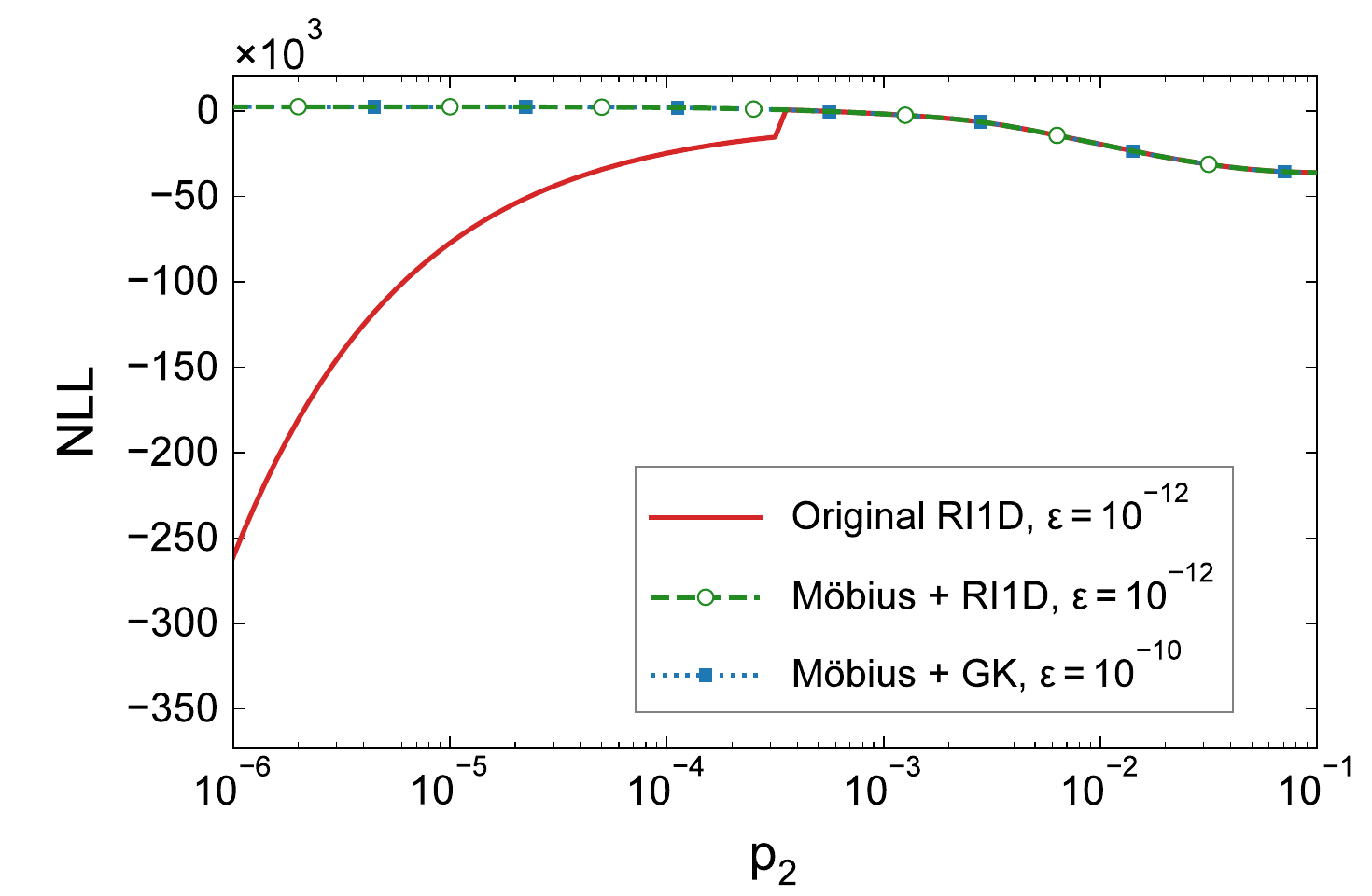}
        \par\vspace{1mm}
        (b)

    \end{minipage}
    \hfill
    \begin{minipage}{0.32\linewidth}
        \centering
        \includegraphics[width=\linewidth]
        {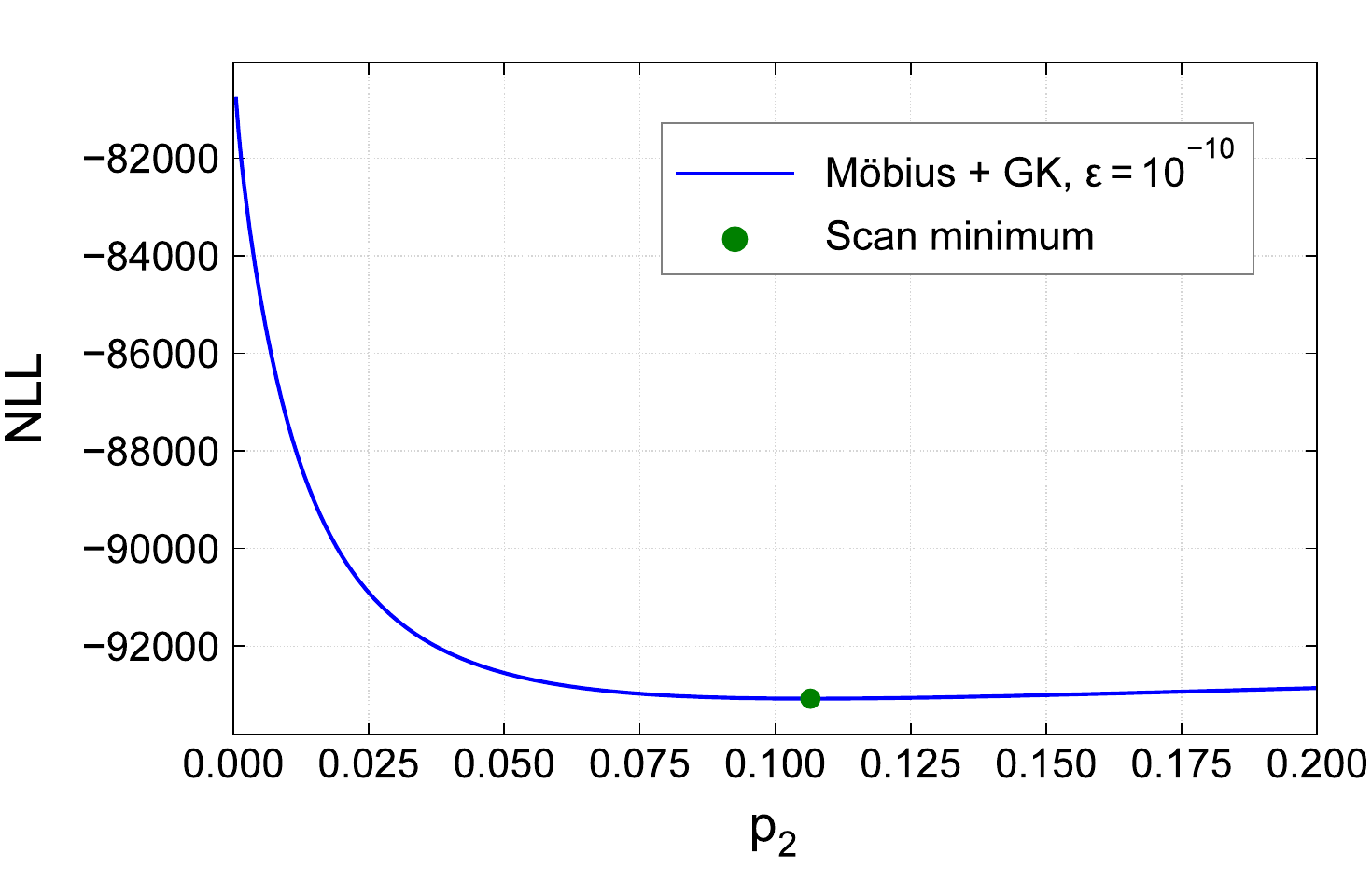}
        \par\vspace{1mm}
        (c)

    \end{minipage}
    \caption{
    Improved evaluation of the NRSPS normalization and the likelihood using the M\"obius variable transformation.
    (a)~NRSPS normalization integral as a function of $p_2$, comparing the original RooIntegrator1D (RI1D) with $\varepsilon=10^{-12}$, M\"obius-transformed RI1D with $\varepsilon=10^{-12}$, and M\"obius-transformed adaptive Gauss--Kronrod integration with $\varepsilon=10^{-10}$.
    (b)~Corresponding NLL scans for $p_2$ using the same three integration configurations, with the other 16 fit parameters fixed at the values returned by the non-converged fit using the original RI1D at $\varepsilon=10^{-12}$ (the runaway case in Tab.~\ref{tab:precision_test}).
    (c)~NLL scan for $p_2$ around the corrected best-fit solution using M\"obius-transformed adaptive Gauss--Kronrod integration with $\varepsilon=10^{-10}$, with the other 16 parameters fixed at their corrected best-fit values.
    The green marker indicates the minimum of the scan.
    The variable transformation removes the severe underestimation of the normalization integral and the associated anomalously low NLL values observed with the original integration method.
    }
    \label{fig:solution_restore}
\end{figure}

The integration variable $m$ is transformed to $t\in[0,1]$ through
\begin{equation}
m(t)=m_{\mathrm{th}}+\frac{Lt}{\lambda+(1-\lambda)t},
\label{eq:mobius_map}
\end{equation}
where $m_{\mathrm{th}}=2m_{J/\psi}$ is the $J/\psi J/\psi$ threshold, $L=m_{\max}-m_{\mathrm{th}}$ with $m_{\max}=15~\mathrm{GeV}$ and $\lambda=L/p_2$. This choice increases the sampling density near threshold as $p_2$ decreases.
The NRSPS normalization integral then becomes
\begin{equation}
I_{\mathrm{NRSPS}}=\int_0^1 f_{\mathrm{NRSPS}}(m(t))\frac{L\lambda}{[\lambda+(1-\lambda)t]^2}\,dt.
\label{eq:mobius_int}
\end{equation}
For small $p_2$, the transformation concentrates sampling points near threshold, where the NRSPS contribution is concentrated.
Figure~\ref{fig:solution_restore}(a) compares the NRSPS normalization integrals obtained using the original RooIntegrator1D with $\varepsilon=10^{-12}$, M\"obius-transformed RooIntegrator1D with $\varepsilon=10^{-12}$, and M\"obius-transformed adaptive Gauss--Kronrod integration with $\varepsilon=10^{-10}$. At $p_2=4.33\times10^{-6}$, the value returned by the non-converged fit (Tab.~\ref{tab:precision_test} runaway case), the M\"obius-transformed calculation gives $I_{\mathrm{NRSPS}}\approx2.77\times10^{-8}$, compared with $3.05\times10^{-100}$ from the original RooIntegrator1D. The two transformed integration methods give consistent results across the scanned $p_2$ range, supporting the reliability of the numerical integration with suitable variable transformation and eliminating the severe underestimation observed with the original method. The corresponding NLL scans in Fig.~\ref{fig:solution_restore}(b) are also smooth after the transformation, without the anomalously low NLL values at small $p_2$.

The full 17-parameter fit is then repeated from the same initial parameter values in the runaway case, using M\"obius-transformed adaptive Gauss--Kronrod integration with $\varepsilon=10^{-10}$ for the NRSPS normalization. The pseudo-dataset and all other aspects of the fit remain unchanged. The fit converges with $\mathrm{minNLL}\approx-93079.8$ and $p_2=0.107\pm0.010$, consistent with the reference solution in Tab.~\ref{tab:precision_test}. The corresponding NLL scan for $p_2$ exhibits a smooth minimum, as shown in Fig.~\ref{fig:solution_restore}(c). As an independent check, we also test the alternative transformation $m=m_{\mathrm{th}}+p_2z^2$, where $z\in[0,\sqrt{L/p_2}]$, a simple quadratic map that likewise concentrates sampling near threshold. Using adaptive Gauss--Kronrod integration with $\varepsilon=10^{-10}$, this transformation also recovers the reference likelihood minimum. These results show that improving the numerical sampling near threshold can stabilize the NRSPS normalization and restore convergence of the full likelihood fit without modifying the underlying mass-spectrum model.

\section{Robustness and Discussion}
\label{sec:discussion}

The two numerical failures examined in Secs.~\ref{sec:discontinuities} and~\ref{sec:runaway} illustrate how inaccurate numerical evaluation of the PDF normalization can distort the likelihood. In the first case, the integrator's decision to stop or continue can change between neighboring parameter points, producing discontinuities in the normalization and hence the NLL, even when the underlying physical density is smooth, which can stall MIGRAD at a spuriously low NLL caused by the incorrect numerical integral. In the second case, the NRSPS distribution becomes increasingly concentrated near the $J/\psi J/\psi$ threshold as $p_2$ decreases. The integration grid may then fail to resolve the narrow peak, leading to a severely underestimated normalization and an artificially low-NLL region into which the minimization can be drawn.

To evaluate the robustness of the improved numerical integration strategies discussed above---tightening the integration tolerances, adopting an alternative adaptive integration algorithm (using Gauss--Kronrod integration as an example), and applying a suitable variable transformation (as demonstrated using the M\"obius transformation for the NRSPS integral in Sec.~\ref{sec:solution})---we repeat the 500 fits for each of the three strategies, using the same randomly generated initial parameter values as in the original study (Fig.~\ref{fig:minnll_scan500}).
The performance of the three strategies is summarized in Tab.~\ref{tab:integrator_robustness}, including their effectiveness in recovering anomalously low NLL values and improving fit convergence.
Tightening the integration tolerances in RooIntegrator1D from $\varepsilon=10^{-7}$ to $10^{-8}$, $10^{-10}$, or $10^{-12}$ recovers nonnegative $\Delta\mathrm{NLL}$ values for $86\%$ of the 69 fits originally exhibiting anomalously low NLL values with $\Delta\mathrm{NLL}<0$, where $\Delta\mathrm{NLL}=\mathrm{NLL}-\mathrm{NLL}_G$ and $G$ denotes the reference global minimum. Among the 390 fits that did not reach $G$ with RooIntegrator1D at $\varepsilon=10^{-7}$, $54\%$ satisfy the convergence criteria and $53\%$ converge to $G$ with these tighter tolerances. Adaptive Gauss--Kronrod integration at $\varepsilon=10^{-7}$ yields the corresponding anomalous-NLL recovery rate of $96\%$, the convergence rate of $81\%$, and the fraction converging to $G$ of $78\%$. Further applying the M\"obius transformation to the NRSPS integral with adaptive Gauss--Kronrod eliminates the remaining anomalously low-NLL cases and achieves convergence for $18\%$ of the fits that did not converge with adaptive Gauss--Kronrod integration alone, with $14\%$ of those that did not reach $G$ under adaptive Gauss--Kronrod integration now converging to $G$. Combining the results from the three strategies eliminates all anomalously low NLL values and achieves an overall convergence rate of $87\%$ among the 390 fits that did not converge to $G$ with the original configuration, with $84\%$ converging to $G$.

\begin{table}[htbp]
\centering
\caption{
    Robustness of three numerical integration strategies using the same 500 sets of randomly generated initial parameter values as in Fig.~\ref{fig:minnll_scan500}. The strategies include tightening the integration tolerances with RooIntegrator1D (RI1D) at $\varepsilon=10^{-8},10^{-10},10^{-12}$, adopting adaptive Gauss--Kronrod (adaptive GK) integration at $\varepsilon=10^{-7}$, and applying the M\"obius transformation to the NRSPS integral with adaptive GK integration at $\varepsilon=10^{-7}$. These strategies are evaluated on the 390 fits that did not converge to the reference global minimum ($G$) using RI1D at $\varepsilon=10^{-7}$.
    The three performance metrics are defined as follows. ``Anomalous-NLL recovery'' denotes the fraction of the 69 original fits with $\Delta\mathrm{NLL}<0$ for which $\Delta\mathrm{NLL}\geq0$ is recovered, where $\Delta\mathrm{NLL}=\mathrm{NLL}-\mathrm{NLL}_G$ and $\mathrm{NLL}_G$ is the NLL at $G$. ``Converged'' requires status $= 0$, covQual $= 3$, and EDM $<10^{-3}$, while ``Converged to $G$'' additionally requires reaching the reference minimum.
    Each entry is given as $(N_{\mathrm{ok}}/N_{\mathrm{test}})$~(\%), where $N_{\mathrm{ok}}$ and $N_{\mathrm{test}}$ denote the numbers of fits satisfying the corresponding criterion and considered for that metric, respectively, with the percentage given in parentheses.
    The M\"obius transformation is applied only to cases unresolved by adaptive GK, where $N_{\mathrm{test}}$ counts the fits still failing that metric after adaptive GK. ``Combined (best result per fit)'' represents the combined performance, selecting one result per fit by prioritizing convergence to $G$, followed by convergence and smallest nonnegative $\Delta\mathrm{NLL}$.
    }
    \label{tab:integrator_robustness}
\footnotesize
\setlength{\tabcolsep}{4pt}
    \begin{tabular}{lccc}
\hline
        Method & Anomalous-NLL recovery & Converged & Converged to $G$ \\
        & $(N_{\mathrm{ok}}/N_{\mathrm{test}})$~(\%) & $(N_{\mathrm{ok}}/N_{\mathrm{test}})$~(\%) & $(N_{\mathrm{ok}}/N_{\mathrm{test}})$~(\%) \\
\hline
        Default RI1D $\varepsilon=10^{-7}$ & 0/69~(0\%) & 112/500~(22\%) & 110/500~(22\%) \\
\hline
        Tighter RI1D & 59/69~(86\%) & 212/390~(54\%) & 206/390~(53\%) \\
        Adaptive GK $\varepsilon=10^{-7}$ & 66/69~(96\%) & 314/390~(81\%) & 306/390~(78\%) \\
        M\"obius NRSPS + adaptive GK $\varepsilon=10^{-7}$ & 3/3~(100\%) & 14/76~(18\%) & 12/84~(14\%) \\
\hline
        Combined (best result per fit) & 69/69~(100\%) & 339/390~(87\%) & 327/390~(84\%) \\
\hline
\end{tabular}
\end{table}

\begin{figure}[htbp]
        \centering
        \begin{tikzpicture}
            \node[inner sep=0] (fig)
        {\includegraphics[width=0.96\linewidth]
        {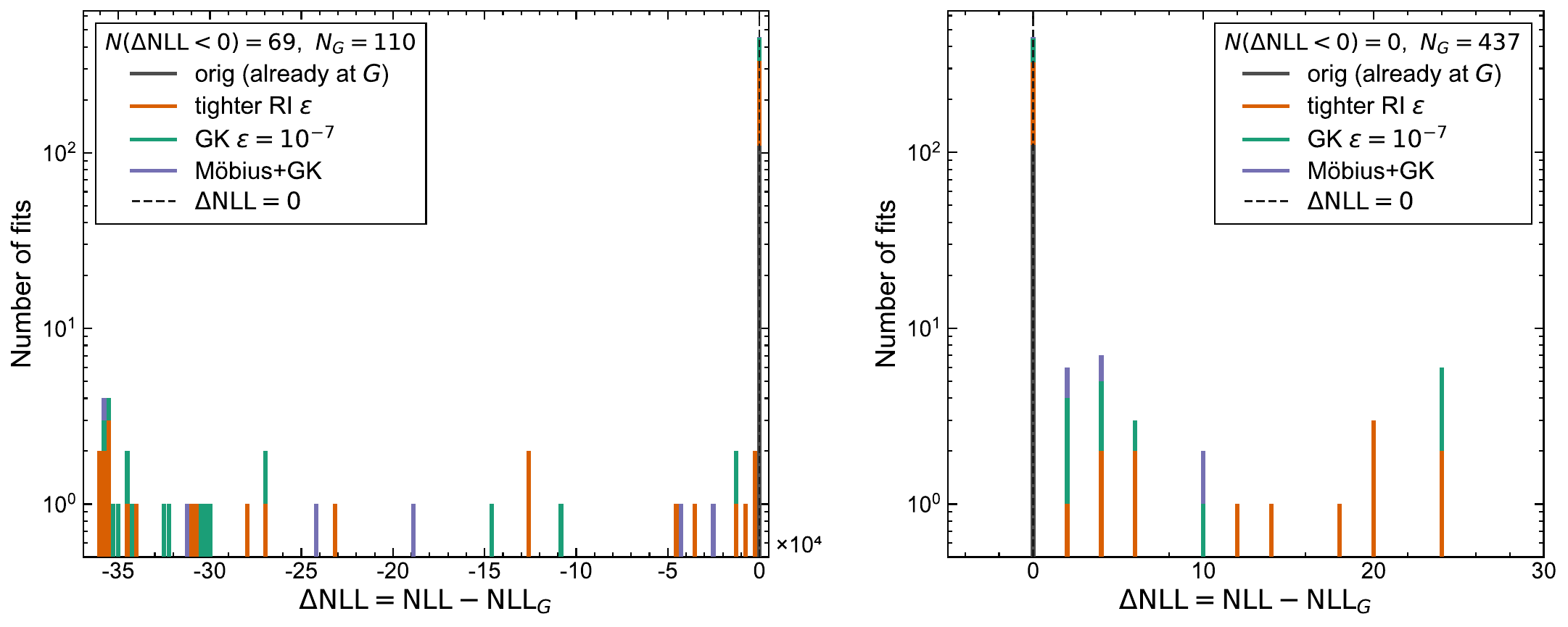}};
            \coordinate (xL) at ($(fig.south west)!0.02!(fig.south east)$);
        \coordinate (xR) at ($(fig.south west)!0.56!(fig.south east)$);
        \coordinate (yT) at ($(fig.south west)!0.96!(fig.north west)$);
            \node[anchor=north west, fill=white, inner sep=1pt]
        at (xL|-yT) {\textbf{(a)}};
            \node[anchor=north west, fill=white, inner sep=1pt]
        at (xR|-yT) {\textbf{(b)}};
        \end{tikzpicture}
    \caption{
    Comparison of the $\Delta\mathrm{NLL}$ distributions for the 500 fits obtained with (a)~the original RooIntegrator1D (RI1D) at $\varepsilon=10^{-7}$ and (b)~the selected results from the three improved numerical integration strategies: tighter RI1D tolerances with $\varepsilon=10^{-8}/10^{-10}/10^{-12}$ (orange), adaptive Gauss--Kronrod (GK) integration (green), and M\"obius-transformed GK integration of the NRSPS integral (purple).
    Here, $\Delta\mathrm{NLL}=\mathrm{NLL}_{\mathrm{fit}}-\mathrm{NLL}_G$, where $\mathrm{NLL}_G$ denotes the NLL at the reference global minimum $G$ (black dashed). The annotations $N(\Delta\mathrm{NLL}<0)$ and $N_G$ indicate the numbers of fits with negative $\Delta\mathrm{NLL}$ and those converging to $G$, respectively.
    For each fit, the $\Delta\mathrm{NLL}$ value shown in panel~(b) is taken from the selected result, prioritizing convergence to $G$, followed by convergence, and finally the smallest nonnegative $\Delta\mathrm{NLL}$. Fits already converged to $G$ with the original method are shown in gray. The histograms are stacked with different colors indicating the selected strategies, with each fit assigned the same color in both panels.
    }
    \label{fig:nll_three_family}
\end{figure}

Figure~\ref{fig:nll_three_family} further compares the original $\Delta\mathrm{NLL}$ distribution with that obtained by selecting one result for each of the 500 initial-parameter sets from the three strategies. For each set, the fit results obtained using tighter RooIntegrator1D tolerances, adaptive Gauss--Kronrod integration, and M\"obius-transformed adaptive Gauss--Kronrod integration are compared. Priority is given to results converging to the reference minimum $G$. If no result reaches $G$, a result satisfying the convergence criteria is selected. Otherwise, the result with the smallest nonnegative $\Delta\mathrm{NLL}$ is retained.
The histograms are stacked with different colors indicating the selected strategies, with each fit assigned the same color in both panels. As shown in Fig.~\ref{fig:nll_three_family}, the resulting distribution no longer contains the anomalously low NLL values below the reference global minimum. Together with the individual recovery rates in Tab.~\ref{tab:integrator_robustness}, these results demonstrate that the numerical integration strategies examined here effectively suppress spurious low-NLL solutions arising from numerical integration errors and improve the robustness of the fitting procedure.

The remaining convergence failures are associated with the strong correlations among fit parameters, local minima, and numerical integration errors. Strong parameter correlations can allow changes in one parameter to be compensated by changes in others with little variation in the NLL, while numerical integration errors can affect the estimation of gradients and curvature. These effects can hinder gradient-based minimization, which may become trapped in local minima or fail to converge.
A search using the SIMPLEX algorithm in Minuit~\cite{James:1975dr} can explore alternative regions of the parameter space by comparing NLL values at different points, without relying on gradient or curvature estimates.
This makes the search less sensitive to unreliable local derivative estimates and can provide a better starting point for subsequent refinement with MIGRAD.
Additional improvements can be achieved by adjusting parameter bounds to limit excursions along weakly constrained directions, where large changes in parameter values have little effect on the likelihood, and by reparameterizing strongly correlated parameters to reduce disparities in curvature across parameter directions, thereby improving the stability of the minimization.
With these additional treatments, all 500 fits were successfully brought to convergence at the reference global minimum.

Although demonstrated using a specific pseudo-dataset and Minuit, the numerical difficulties identified here reflect general challenges in likelihood minimization. Extended unbinned likelihood fits relying on numerical PDF normalization, particularly for distributions that can become sharply peaked, may encounter similar problems when integration grids fail to resolve important contributions or when stopping decisions change.
Practical safeguards include examining the likelihood and component normalization integrals along sensitive parameters, adjusting numerical integration tolerances, comparing independent integration algorithms, monitoring shape parameters whose values can produce sharply localized features, and applying suitable integration variable transformations where necessary.

These safeguards, however, come with practical trade-offs. Tighter integration tolerances can substantially increase computational cost. For a representative set of $p_2$ values, the average evaluation time of the NRSPS normalization integral using RooIntegrator1D increases by factors of approximately $70$ and $260$ when the tolerance is reduced from $10^{-7}$ to $10^{-10}$ and $10^{-12}$, respectively. As normalization integrals are evaluated repeatedly during minimization, this additional cost can become substantial. Furthermore, tighter tolerances may offer little improvement in accuracy for numerically stable fits and do not necessarily resolve integration errors when sharply localized features are insufficiently sampled by the integration algorithm. The effectiveness of alternative integration algorithms and variable transformations also depends on the integrand and parameter region, while convergence difficulties associated with strong parameter correlations or local minima may require separate treatments. We therefore recommend retaining the default settings when numerical stability has been verified and applying more stringent tolerances or alternative integration methods when numerical instabilities are identified. Regardless of the numerical settings used, an anomalously low NLL accompanied by failed convergence diagnostics should be treated as a potential numerical warning rather than evidence for a better physical solution.

\section{Conclusions}
\label{sec:conclusions}

Anomalously low NLL values have been investigated using a pseudo-dataset based on the CMS $J/\psi J/\psi$ analysis~\cite{CMS:2022pas,CMS:2023owd,CMS:2025fpt}. Numerical errors in PDF normalization can introduce artificial structures in the likelihood surface and lead to unreliable minimization results. In particular, changes in the integrator's stopping decision as the fit parameters vary can produce discontinuities in the returned normalization, even at fixed integration tolerance. The problem becomes particularly severe when the PDF develops a sharply localized structure that is poorly resolved by the integration grid, allowing the numerical error estimate to indicate convergence despite a substantial normalization error. Such errors can cause the minimization algorithm (in our example, MIGRAD~\cite{James:1975dr}) to encounter and retain parameter configurations with spuriously low NLL values.

Tightening the integration tolerance or using alternative adaptive numerical integration can mitigate these instabilities, but these approaches do not necessarily resolve sharply localized contributions. When evaluating PDF normalizations, an appropriate variable transformation that concentrates the sampling points in the critically sampled region can stabilize the numerical integral and restore reliable minimization. Across the 500 fits studied, these strategies eliminate all anomalously low NLL values and substantially improve convergence to the reference global minimum (Tab.~\ref{tab:integrator_robustness}). These results imply that an apparently lower NLL need not signal a better description of the data. It may instead arise from an inaccurate PDF normalization, even when the numerical integrator reports that its stopping criterion has been satisfied. In particular, non-converged fits that report unusually low NLL values should not be regarded as mathematically preferred solutions. Care is therefore required when interpreting such outcomes in fits that rely on numerically evaluated PDF normalizations.

\section*{Acknowledgements}
This work was supported by the National Natural Science Foundation of China under Grant No.~12535004.

\clearpage

\bibliographystyle{spphys}
\bibliography{references}

\end{document}